%% file: SQuIGGLE-PaA.tex
\documentclass[twocolumn]{aastex701}

\newcommand{\squiggle}{SQuIGG$\vec{L}$E \,}
\newcommand{\squigglecomma}{SQuIGG$\vec{L}$E}

\newcommand{\logM}{log($M_\star/M_\odot$) \,}

\newcommand{\sersic}{S\'{e}rsic \,}

\usepackage{amsmath} 

\usepackage[encapsulated]{CJK}

\usepackage{rotating}

\usepackage{float}
\usepackage{textcomp}
\usepackage{makecell}

\begin{document}

\date{\today}

\submitjournal{ApJ}

\shorttitle{Pa$\alpha$ SFRs of CO luminous \squiggle galaxies}

\shortauthors{Setton et al.}

\title{SQuIGG$\vec{L}$E: Spatially resolved NIRSpec Pa$\alpha$ reveals that the most CO-luminous, $z\sim0.7$ post-starburst galaxies exhibit falling---but still active---star formation behind optically thick dust}

\author[0000-0003-4075-7393]{David J. Setton}\thanks{Email: davidsetton@jhu.edu}\thanks{NHFP Hubble Fellow}
\affiliation{The William H. Miller III Department of Physics and Astronomy, The Johns Hopkins University, Baltimore, MD 21218}
\affiliation{Department of Astrophysical Sciences, Princeton University, 4 Ivy Lane, Princeton, NJ 08544, USA}
\email{davidsetton@jhu.edu}

\author[0000-0002-5612-3427]{Jenny E. Greene}
\affiliation{Department of Astrophysical Sciences, Princeton University, 4 Ivy Lane, Princeton, NJ 08544, USA}
\email{jgreene@astro.princeton.edu}

\author[0000-0001-5063-8254]{Rachel Bezanson}
\affiliation{Department of Physics and Astronomy and PITT PACC, University of Pittsburgh, Pittsburgh, PA 15260, USA}
\email{rachel.bezanson@pitt.edu}

\author[0000-0002-1714-1905]{Katherine A. Suess}
\affiliation{Department for Astrophysical \& Planetary Science, University of Colorado, Boulder, CO 80309, USA}
\email{suess@colorado.edu}

\author[0000-0003-3256-5615]{Justin~S.~Spilker}
\affiliation{Department of Physics and Astronomy and George P. and Cynthia Woods Mitchell Institute for Fundamental Physics and Astronomy, Texas A\&M University, 4242 TAMU, College Station, TX 77843-4242, US}
\email{jspilker@tamu.edu}

\author[0000-0002-1759-6205]{Vincenzo~R.~D'Onofrio}
\affiliation{Department of Physics and Astronomy and George P. and Cynthia Woods Mitchell Institute for Fundamental Physics and Astronomy, Texas A\&M University, 4242 TAMU, College Station, TX 77843-4242, US}
\email{donofr19@tamu.edu}

\author[0000-0002-0766-1704]{Elia Cenci}
\affiliation{Department of Astronomy, University of Geneva, Chemin Pegasi 51, Versoix CH-1290, Switzerland}
\email{elia.cenci@unige.ch}

\author[0000-0002-1109-1919]{Robert~Feldmann}
\affiliation{Department of Astrophysics, Universit\"at Z\"urich, Winterthurerstrasse 190, Zurich CH-8057, Switzerland}
\email{feldmann@physik.uzh.ch}

\author[0009-0009-9810-3334]{Alex~J.~Geiger}
\affiliation{Department of Physics and Astronomy and George P. and Cynthia Woods Mitchell Institute for Fundamental Physics and Astronomy, Texas A\&M University, 4242 TAMU, College Station, TX 77843-4242, US}
\email{ageiger@tamu.edu}

\author[0000-0003-4700-663X]{Andy~D.~Goulding}
\affiliation{Department of Astrophysical Sciences, Princeton University, 4 Ivy Lane, Princeton, NJ 08544, USA}
\email{goulding@astro.princeton.edu}

\author[0000-0002-7613-9872]{Mariska~Kriek}
\affiliation{Leiden Observatory, Leiden University, P.O. Box 9513, 2300 RA Leiden, The Netherlands}
\email{kriek@strw.leidenuniv.nl}

\author[0009-0005-4226-0964]{Anika Kumar} 
\affiliation{Laboratory for Multiwavelength Astrophysics, School of Physics and Astronomy, Rochester Institute of Technology, 84 Lomb Memorial Drive, Rochester, NY 14623, USA} 
\email{ak8532@rit.edu}

\author[0000-0002-0696-6952]{Yuanze~Luo}
\affiliation{Department of Physics and Astronomy and George P. and Cynthia Woods Mitchell Institute for Fundamental Physics and Astronomy, Texas A\&M University, 4242 TAMU, College Station, TX 77843-4242, US}
\email{yluo37@tamu.edu}

\author[0000-0002-7064-4309]{Desika~Narayanan}
\affiliation{Department of Astronomy, University of Florida, 211 Bryant Space Science Center, Gainesville, FL 32611, USA}
\email{desika.narayanan@ufl.edu}

\author[0000-0003-1535-4277]{Margaret E. Verrico}
\affiliation{University of Illinois Urbana-Champaign Department of Astronomy, University of Illinois, 1002 W. Green St., Urbana, IL 61801, USA}
\affiliation{Center for AstroPhysical Surveys, National Center for Supercomputing Applications, 1205 West Clark Street, Urbana, IL 61801, USA}
\email{verrico2@illinois.edu}

\author[0000-0002-6768-8335]{Pengpei~Zhu}
\affiliation{Cosmic Dawn Center (DAWN), Denmark}
\affiliation{DTU Space, Technical University of Denmark, Elektrovej 327, 2800 Kgs. Lyngby, Denmark}
\email{penzhu@space.dtu.dk}

\submitjournal{ApJ}

\begin{abstract}

One of the fundamental uncertainties in the study of galaxies at the interface between star formation and quiescence--post-starburst galaxies--is their instantaneous star formation rate. While these galaxies appear quiescent in the rest UV-optical, many of the youngest systems are highly CO luminous, implying dense molecular gas, and potentially star formation. However, their instantaneous star formation rates have proved elusive, as measurements with rest-optical emission lines or UV-to-IR SED fitting yield order-of-magnitude uncertainty. Here, we present JWST/NIRSpec G395M/170LP integral field spectroscopy of the three most CO-luminous massive galaxies in the SQuIGG$\vec{L}$E survey, spatially resolving Pa$\alpha$/Br$\gamma$/Br$\beta$ to map the dust-corrected star formation rate. We find that these galaxies host star formation rates of $27-80$ $M_\odot \ \mathrm{yr^{-1}}$, though $\sim30-40\%$ of the emission may arise from non–star-forming ionization (AGN or shocks), implied by high [FeII]/Pa$\alpha$ in central spaxels. We also find effective $A_{\mathrm{Pa}\alpha}\sim1$ and individual $\sim700$ pc spaxels with $A_{\mathrm{Pa\alpha}}\sim2$, implying significant dust obscuration of nebular regions that contribute little to the rest-optical lines. These galaxies are not consistent with total quiescence; the measured star formation rates place these galaxies on or above the main sequence. However, their star formation rates \textit{are} significantly lower ($\sim0.5-1$ dex) than the peak star formation rates inferred from spectrophotometric fitting. We propose that the CO luminous post-starburst phase represents the last gasp of star formation in galaxies on the rapid quenching pathway, and that dust-obscured star formation, outflows, and (speculatively) an uncertain CO-to-H2 conversion reconcile their gas exhaustion with observed timescales of fading.
\end{abstract}



\section{Introduction} \label{sec:intro}

One of the central questions in the formation of massive galaxies is the physical process by which the compact cores of massive, quiescent galaxies, assemble. Archaeological studies of the centers of the most massive galaxies (\logM$>10^{11} M_\odot$) in the local Universe have revealed that mass buildup occurs on very short ($\lesssim 1$ Gyr) timescales with peak star formation rates of hundreds of solar masses per year, followed by rapid shutdown \citep{Thomas2005, McDermid2015, FerreMateu2017}. At high-z, direct studies, particularly with the new James Webb Space Telescope (JWST), have detected the progenitors of such systems in the early universe, with the presence of mature massive galaxies at and above cosmic noon corroborating these extreme star formation rates and timescales \citep[e.g.,][]{VanDokkum2015,Kriek2016, Glazebrook2017, Glazebrook2024, Man2021, Carnall2019, Carnall2023b, Carnall2024, Beverage2024, Beverage2025, Antwi-Danso2025, McConachie2025, Zhang2025_QG_numberdensity, Zhang2026_QG_numberdensity_runback}. 

While it is clear that such rapid formation is the dominant production pathway for massive galaxies at $z\gtrsim1$ \citep{Rowlands2015, Wild2016, Zick2018, Belli2019, Setton2023,Park2024_numberdensity}, a clear physical picture of what triggers these bursts and how they subsequently remove or exhaust their molecular gas has remained elusive. A canonical picture of major-merger induced gas inflows that enhance star formation and active galactic nucleus (AGN) activity has emerged \citep[e.g.,][]{Hopkins2008a}. However, direct observational evidence linking the stages of this timeline has proved difficult to observe in galaxies actively undergoing this process at high-z, as even JWST struggles to build large samples of galaxies at each stage of this process, let alone to resolve all phases of their stars and gas.

It is for this reason that post-starburst galaxies \citep{Dressler1983, Zabludoff1996} have emerged as the eminent laboratory for studying the physics of rapid quenching. Post-starburst galaxies are characterized by their A type dominated rest-optical spectra, which indicate that they are dominated in stellar mass by a recent period of star formation that rapidly shut off, and as such, the most massive post-starburst galaxies (with $\log(M_\star/M_\odot)>11$) at lower redshift may serve as analogues for the rapid formation channel that dominates at high-redshift. While massive post-starburst galaxies are very rare in the local universe \citep{Goto2003_numberdensity, Quintero2004, Wild2009, Pattarakijwanich2016}, their number density rises steeply at intermediate-redshift \citep[$z\sim0.5-2$, e.g.,][]{Whitaker2012a, Wild2016, Rowlands2018a, Belli2019, Gobat2020_gas_fraction_number_density,Setton2023, Park2024_numberdensity, Adscheid2025}. This intermediate redshift regime has been rich for multi-wavelength study with facilities like ALMA and NOEMA, which have quantified the dust and gas population in the wake of rapid quenching.

A surprising result of these studies has been that, in contrast with massive quiescent systems that tend to be quite depleted in gas and dust \citep{Saintonge2011, Young2011, Sargent2015, Davis2016, Spilker2018,Michalowski2019, Whitaker2021b, Williams2021, Michalowski2024, Adscheid2025, Suess2025}, many post-starburst galaxies are CO- or dust continuum-luminous \citep{French2015, French2018a, Rowlands2015, Alatalo2016b, Suess2017, Yesuf2017, Smercina2018, Belli2021, Bezanson2022a, Woodrum2022, Otter2022, Baron2023, Wu2023, Michalowski2024, Umehata2025, Suess2025, Spilker2025}, suggesting the presence of significant interstellar medium (ISM) reservoirs, with varying works finding pronounced trends that suggest that massive recently quenched and post-merger systems are particularly likely to retain their ISM. One possible interpretation of this result is that the final removal of molecular gas occurs \textit{after} the cessation of star formation \citep[e.g.,][]{Bezanson2022a}, necessitating mechanical gas removal via outflows \citep[e.g.,][Luo et al. 2026 submitted]{Smercina2022, Belli2023, French2023a, Valentino2025}, tidal stripping \citep[e.g.,][]{Spilker2022, Donofrio2025, Broadbelt2026}, or some combination of the two to align post-starburst galaxies with their quiescent descendants.

However, an alternative explanation for the retention of ISM in these post-starburst systems is that they are being mis-identified as quenched. Many post-starburst galaxies are IR luminous \citep[e.g.,][]{Alatalo2017, Yesuf2017, Smercina2018, Baron2022a, Baron2023, Setton2025_squiggle, Baron2026_sample, Baron2026_SED}, necessitating a source of luminosity that is coupled to the ISM to produce the observed luminous hot and warm dust. In the extreme interpretation, these high IR (and sometimes, radio) luminosities can imply that these galaxies are still at the peak of their star formation, and that dust geometry effects mask the emission from O and B stars entirely in the rest-optical \citep{Smail1999, Poggianti2000, Baron2022a, Cenci2025_PSB}, implying that these galaxies are not at all ``post"-starburst. However, the longer timescales associated with these IR tracers produce significant degeneracy in the precise fraction of the IR luminosity dominated by ``instantaneous" ($\sim10$ Myr) star formation and long timescale ($\sim100$ Myr) star formation \citep{Hayward2014, Utomo2014, Leja2019, Baron2023, Wild2025, Setton2025_squiggle}, especially given the uncertainty in how the dust geometry differs around different aged populations (an effect which is typically accounted for in SED fits with a simple ``two screen" \citealt{Charlot2000} geometry). The potential presence of AGN dust heating, particularly in the mid-IR \citep[e.g.,][]{Ciesla2015, Salim2016, Leja2018}, presents another degree of freedom that muddies the interpretation of the instantaneous star formation rate. Altogether, this leaves room for a wide range of degenerate solutions in the most IR luminous, CO-luminous post-starburst galaxies, ranging from actively starbursting to truly quiescent \citep{Setton2025_squiggle}.

Ideally, one could address this degeneracy by directly accessing instantaneous tracers of star formation such as Hydrogen recombination lines. However, at the proposed optical depths of the most centrally concentrated star forming regions of post-starburst galaxies, even Balmer decrement-corrected H$\alpha$/H$\beta$ emission would be insensitive to the mostly deeply embedded star formation \citep{Baron2023}, providing room for even H$\alpha$ that is only weakly detected in massive post-starburst galaxies \citep{Zhu2025} to be consistent with star formation rates of $\sim100s$ of solar masses-per-year in the most-extreme geometric configurations. Some work with low redshift post-starburst galaxies has utilized longer wavelength emission lines to measure lower star formation rates than those implied by the IR luminosity \citep{Smercina2018, Luo2022, French2023a}, suggesting that this is not the case, but the lack of observing facilities that can access such lines at intermediate-redshift where massive post-starburst samples are abundant has made it impossible to obtain instantaneous measurements for the most CO luminous and massive post-starburst galaxies. 

The advent of JWST has changed this landscape. The NIRSpec instrument is not just capable of detecting near-IR recombination lines such as Paschen and Brackett series emission that are pristine, dust-insensitive tracers of the instantaneous star formation rate at intermediate-z; it is also capable of spatially resolving them on sub-kpc scales \citep[e.g.,][]{Neufeld2024_PaA, Wozniak2026_FRESCO_PaA, Liu2026_PaA, Alberts_2026_midIR_SFR, Vanicek2026_PaA_BrA}. In massive, CO luminous post-starburst galaxies, NIRSpec IFU observations have the potential to penetrate central dust screens and obtain attenuation-corrected Pa$\alpha$ luminosities to obtain secure measurements of the instantaneous star formation rate for the first time. Here, we present NIRSpec/IFU observations of such galaxies, using the three most CO luminous galaxies from the \squiggle Survey of CO $z\sim0.7$ post-starburst galaxies \citep{Suess2022a, Bezanson2022a, Setton2025_squiggle} as an extreme test case of buried star formation. Using Pa$\alpha$/Br$\gamma$/Br$\beta$, we perform a spatially-resolved dust correction to measure the star formation rate in these systems, and, in conjunction with star formation histories measured using UV-to-FIR spectrophotometric data, place these galaxies in the context of the sequence from starburst to quiescence.

This paper is laid out as follows. In Section \ref{sec:data}, we present the \squiggle sample and our reduction of new NIRSpec/IFU, Magellan/FIRE, and ALMA data. In Section \ref{sec:analysis}, we measure integrated Balmer decrements and spatially resolved Pa$\alpha$/Br$\gamma$/Br$\beta$ decrements to obtain the dust-corrected star formation rate and investigate the ionization state of the ISM in these galaxies. In Section \ref{sec:discussion}, we place these galaxies in an evolutionary context relative to co-eval samples of starburst galaxies and quiescent post-starburst galaxies. Finally, in Section \ref{sec:conclusion}, we summarize our results and suggest future avenues for exploring this population with JWST and future planned surveys. Throughout this work, we adopt the best-fit cosmological parameters from the WMAP 9 year results \citep{Hinshaw2013}: $H_0 = 69.32 \ \mathrm{km \ s^{-1} \ Mpc^{-1}}$, $\Omega_m = 0.2865$, and $\Omega_\Lambda = 0.7135$, utilize a Chabrier initial mass function \citep{Chabrier2003}, and quote AB magnitudes. All stellar masses reported as surviving stellar mass after correcting for mass loss. 

\section{Data} \label{sec:data}

\begin{figure*}
    \centering
    \includegraphics[width=\textwidth]{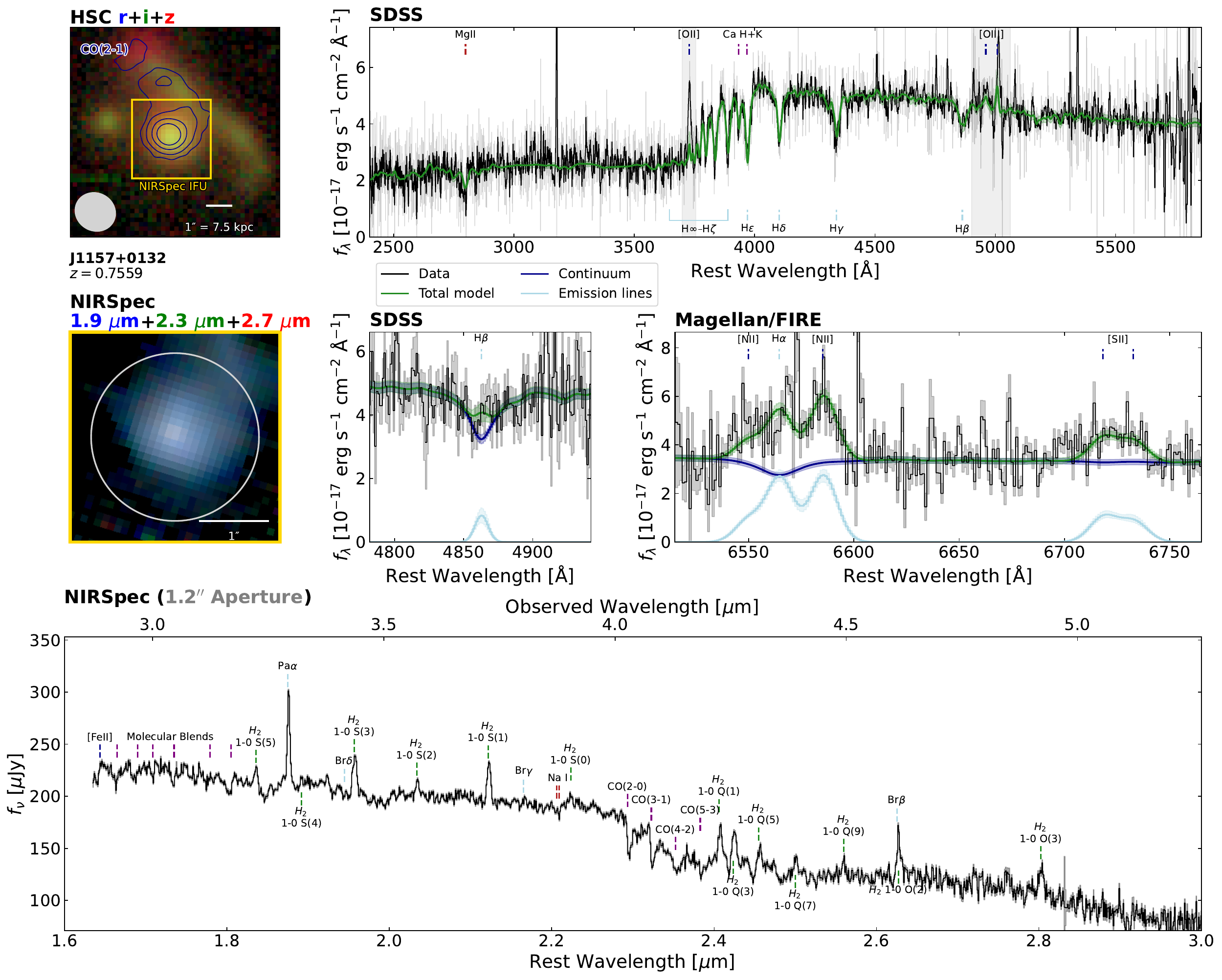}
    \caption{(Top Left): The HSC riz image of J1157+0132, with contours showing the distribution of CO(2-1) flux. The synthesized beam is shown as a gray ellipse. The orange box shows a subset of the NIRSpec IFU field of view, which is shown below in collapsed 1.9/2.3/2.7 $\mu$m color image. In that panel, the 1.2" aperture used for spectral extraction is shown in light blue. (Top right): The rest-optical spectroscopy for this galaxy. On the top, we show the SDSS spectrum, with the best fitting model from \cite{Setton2025_squiggle} shown in green. Below, we show zoom ins of the H$\beta$ complex, with best fitting emission shown in light blue and the continuum only model shown as dark blue. We also show the H$\alpha$/[NII]/[SII] complex as observed by Magellan/FIRE. (Bottom): The NIRSpec IFU G395M spectrum in the collapsed 1.2" aperture. Key emission and absorption lines are labeled, including hydrogen recombination lines (light blue), molecular hydrogen ro-vibrational lines (green), molecular absorption lines (purple), outflow-sensitive ISM features (red), and other density- and temperature-sensitive lines (dark blue). The same data for the other two sources studied in this work are shown in Figures \ref{fig:J0910_fulldata} and \ref{fig:J0907_fulldata}.}
    \label{fig:J1157_fulldata}
\end{figure*}

\subsection{The \squiggle Sample}

\input{tables/source_properties_table}

In this work, we utilize galaxies with multi-wavelength observations that were collected as a part of the \squiggle collaboration. The \squiggle sample was selected from the Sloan Digital Sky Survey Data Release 14 \citep[SDSS DR14,][]{Abolfathi2018}, with the goal of obtaining a representative sample of $z=0.5-0.9$ massive post-starburst galaxies that could be used to study the evolution of this population as a function of various properties. These galaxies were selected using $U_m$, $B_m$, and $V_m$ rest-frame colors defined in \cite{Kriek2010}, in a part of color space that exclusively contains galaxies with strong Balmer breaks and blue slopes redward of the breaks. Galaxies in this part of parameter space are dominated in the rest-optical by A-type stellar populations, indicating a recent decline in their star formation. A full description of the sample is presented in \cite{Suess2022a}. In total, the spectroscopic sample consists of 1318 galaxies, with a median redshift of 0.7 and a median $\log(M_\star/M_\odot) \sim 11.2$.

In this work, we are particularly interested in the most CO luminous galaxies in \squiggle. Several observing campaigns targeted a subset of the \squiggle sample ALMA Band 4 (probing CO(2-1)) to understand their molecular gas properties \citep{Suess2017, Bezanson2022a, Setton2025_squiggle}, resulting in a total sample of 50 galaxies with constraints on $L'_\mathrm{CO(2-1)}$. \cite{Setton2025_squiggle} presented this full sample and showed that, in the most CO luminous galaxies, there exists an order of magnitude uncertainty in their instantaneous ($\sim10$ Myr) star formation rate due to degeneracies in the modeling between the level of attenuation experienced by the youngest stellar populations (the birth cloud attenuation in the two-screen \citealt{Charlot2000} dust model, $\tau_{BC}$) and the amount of star formation in that bin. A wide range of energy balance solutions between dust heating by the youngest and older (10-$\sim$a few hundred Myr) stellar populations can fit the data equally well. As such, even discounting the uncertainty in the molecular gas mass due to systematics in the conversion between the CO luminosity and $M_{H_2}$, the depletion times of these galaxies remain highly uncertain. 

Here, we focus on the three most CO luminous galaxies identified in \cite{Setton2025_squiggle}. We show the properties of these galaxies in Table \ref{tab:source_props} and show their rest-optical imaging and spectroscopy, along with CO(2-1) contours, in Figures \ref{fig:J1157_fulldata}, \ref{fig:J0910_fulldata}, and \ref{fig:J0907_fulldata}. At their extremely high ($>10^{10}$ K km s$^{-1}$ pc$^{2}$) CO(2-1) luminosities, under the CO-to-$H_2$ conversions used in that work they are consistent with $M_{H_2}/M_\star\sim20-50\%$, comparable to star forming galaxies at similar mass and redshift \citep[e.g.,][]{Tacconi2013, Freundlich2019}. As such, they are critical tests of the star formation rate in CO luminous post-starburst galaxies; are they inefficient in their star formation given their vast reservoirs, or are they host to deeply obscured star formation that can reconcile their depletion times with observed trends? In the following sections, we present the new observations that address those questions by spectroscopically targeting instantaneous star formation tracers in the rest-optical and near-IR.

\subsection{Magellan/FIRE Spectroscopy}

We obtained follow-up spectroscopy targeting H$\alpha$ emission in these sources using the Folded-port InfraRed Echellete (FIRE) instrument on the Magellan Baade Telescope. These observations were obtained over a range of nights; we observed J1157+0132 on 03-18-2025, J0910+0218 on 01-08-2023, and J0907+0423 on 02-20-2022. We conducted all observations using an ABBA nodding pattern for NIR difference imaging. All science frames were taken with 900 second integrations, resulting in 1 hour of total integration for each target. J1157+0132 and J0910+0218 were observed with a 0.75" slit, while J0907+0423 was observed with a 1.00" slit.

To reduce the FIRE data, we utilize \texttt{PypeIt}, a python-based spectroscopic reduction pipeline \citep{Prochaska2020_pypeit_code, Prochaska2020_pypeit_paper}. We perform wavelength calibration using the OH sky lines, and perform automated trace finding using the continuum. After extracting 1D spectra from each AB combination and performing a telluric correction and flux calibration using observed stars, we combined the final spectra in a coadd and binned to R=6000, the approximate FIRE resolution. We note that in the case of J0907+0423 (which sits at $z=0.663$, situating the H$\alpha$/[NII]/[SII] in a region heavily populated by sky lines), the sky is particularly bright in two of our frames, rendering them unusable. As such, we only use a single AB combination for this source, resulting in a lower signal-to-noise and a poorly constrained H$\alpha$ non-detection. In the other two sources, H$\alpha$ and both lines of the [NII] and [SII] doublets are clearly detected.

\subsection{JWST/NIRSpec Spectroscopy}

We observed each of our targets in the NIRSpec/IFU \citep{Boker2022_NIRSpec_IFU} G395M/170LP configuration as part of JWST-GO \#6719 (PI: D. Setton), with the primary science goal of detecting and characterizing the spatial distribution of Pa$\alpha$ emission in these sources, with longer wavelength Brackett emission features out to Brackett$ \beta$ also covered by this grating setting to enable a correction for dust attenuation. Each galaxy was observed in an 8 point small cycling dither pattern, with 15 groups/integration and 1 integration/exposure. These settings resulted in a total exposure time of 31.1 minutes/galaxy, and were chosen to achieve a continuum signal to noise of $\sim15$ at observed-frame $\sim$3$\mu$m, the wavelength of Pa$\alpha$ for our targets, while maximizing our sampling of the PSF. The observations were carried out on April 5, 2025 for J0910+0218 and J0907+0423, and May 26, 2025 for J1157+0132. 

We reduce our NIRSpec IFU data using the JWST data reduction pipeline \citep[version 3.0.0,][]{Rigby2023_jwstpipeline, Bushouse2025} using the \texttt{PMAP1591} calibration reference data set. We perform a fairly standard reduction of the data, though we omit any background subtraction steps in favor of a custom background subtraction. In the Level 3 reduction stage, we utilize the \texttt{IFUALIGN} coordinate system to minimize artifacts from interpolating onto a new pixel grid, and use the native 0.1'' pixel grid as the coordinates for our spatial pixels (hereafter referred to as ``spaxels"). Following \cite{Rigby2025_jwstpipeline} and \cite{Khullar2026_LEGGOS}, we perform a custom background subtraction using the predictions from the JWST Backgrounds Tool (JBT) set to the coordinates and time of our observations.

\section{Analysis} \label{sec:analysis}

\subsection{Rest optical line fitting} \label{subsec:FIRE_fitting}

\input{tables/halpha_fit_table_95}

We begin by using integrated measurements of the Balmer series emission to infer the global star formation rate and dust attenuation in a typical manner. To correct line fluxes for the underlying stellar continuum, we rely on the SDSS spectroscopy presented in \cite{Suess2022a}, which covers H$\beta$, and the new Magellan/FIRE spectroscopy that we present here. This measurement requires aperture-matched flux measurements that account for the underlying stellar absorption. To obtain these, we utilize the best fitting continuum spectrum from \cite{Setton2025_squiggle}, and scale the SDSS spectrum to the total model (constrained by SDSS and WISE photometry) using the polynomial calibration from that same fitting (SDSS) and a multiplicative factor (FIRE, but only in the narrow region of H$\alpha$/[NII]/[SII] emission). These aperture corrections are substantial; the SDSS spectra require corrections of 2.2, 1.7, and 2.9 at the wavelengths of H$\beta$ and the FIRE spectra require corrections of 2.8, 2.4, and 4.3 for J1157+0132, J0910+0218, and J0907+0423, respectively. Aperture corrections of this magnitude come with the significant assumption that the fiber/slit being used is covering a region of the galaxy where a uniform scaling to the continuum and line emission can be applied. In later sections, we will show that this assumption is problematic for these galaxies. Nevertheless, we continue by utilizing these values to obtain line fluxes.

We jointly model H$\beta$, H$\alpha$, the [NII] doublet, and (except in the case of J0907+0423 where [SII] is in a region that is unusable due to the telluric) the [SII] doublet. For the continuum, we draw from the posterior of the \cite{Setton2025_squiggle} spectrophotometric fits, correcting the SDSS spectrum using the calibration vector from those fits and the FIRE spectrum in the region of the H$\alpha$ complex with a simple multiplicative factor. All lines are fit with a common velocity dispersion, which is added in quadrature to the instrumental resolution. The results of this fitting are shown in the bottom row of Figure \ref{fig:J1157_fulldata} (and Figures \ref{fig:J0910_fulldata} and \ref{fig:J0907_fulldata}), and the line fluxes for H$\alpha$ and H$\beta$ are presented in Table \ref{tab:halpha}. In J1157+0132, H$\beta$ is only weakly detected in the SDSS spectrum, while the full H$\alpha$/[NII]/[SII] complex is detected. In J0910+0218, all lines, including H$\beta$, are clearly detected. Finally, in J0907+0423, all lines are undetected, leading to very weak limits on the line fluxes, and, subsequently, the attenuation and star formation rate.

We note that in the two galaxies with securely detected H$\alpha$ and [NII], the [NII]/H$\alpha$ ratio is $1.04 \pm^{0.25}_{0.21}$ and $0.62 \pm^{0.12}_{0.11}$, squarely in the composite range of the \cite{Kewley2006} demarcation between star forming and AGN line ratios, further complicating the interpretation of the rest-optical emission lines. In the following section, we will proceed under the assumption that these lines are dominated by star formation, but in Section \ref{subsec:non-sf-ionization}, we will return to the possibility of non-star formation driven ionization and attempt to correct for the presence of AGN, LINER, or shocks.

\subsection{Balmer decrement measurement} \label{subsec:balmer_decrement}

We now proceed with using these lines to measure the star formation rates. \cite{Zhu2025}, who analyzed Keck/NIRES H$\alpha$ spectroscopy of a broader sample of \squiggle galaxies, did not directly use the Balmer decrement to measure the star formation rate, and instead utilized the dust constraints from SED fits to correct the H$\alpha$ for dust. This was done for two reasons. First, in all those sources, H$\beta$ was only marginally, if ever, detected. Second, in the majority of those sources, even H$\alpha$ was only marginally detected. However, in this work, where we are interested in directly constraining the nebular dust attenuation and where 2/3 sources have strong H$\alpha$ detections, we proceed with a direct Balmer decrement measurement, noting again the considerable systematic uncertainty imparted by the aperture corrections.

In all sources, we find that the H$\alpha$/H$\beta$ ratio is highly uncertain, largely due to the degeneracy between the H$\beta$ emission and the strong absorption implied by the A-star dominated stellar population. We utilize the measured H$\alpha$/H$\beta$ ratios under the assumption of Case B recombination (intrinsic H$\alpha$/H$\beta$=2.86) and a \cite{Cardelli1989} dust law. Only in the case of J0910+0218 is there a strong preference for high nebular attenuation, and even so, the posterior for $A_{V}$ spans $\sim2.1-4.8$ due to the uncertainty in the H$\beta$ flux. In the case of J1157+0132, the Balmer decrement is minimally constrained to be in the range 0-3.5. Finally in the case of J0907+0423, where neither line is confidently detected, the nebular attenuation is entirely unconstrained, with $A_{V}<5.3$ at the 95\% level, and there are essentially no useful constraints we can extract from these limits.

We now utilize the posteriors for these galaxies to measure the H$\alpha$ star formation rate measurement under the assumption that the H$\alpha$ ionizing budget is totally dominated by young stellar populations. To do so, we adopt the calibration from \cite{Murphy2011} (noting that its assumption of a \cite{Kroupa2001} IMF produces very similar calibrations to a \cite{Chabrier2003} IMF, see \cite{Speagle2014}):

\begin{equation}
    \mathrm{SFR} \ [M_\odot / \mathrm{yr}] = 5.37 \times 10^{-42} \times L_{\mathrm{H}\alpha} \ [\mathrm{erg/s}]
\end{equation}

Restricting to positive attenuation, we find that the star formation rate in J1157+0132 spans a 95\% confidence interval of 10-157 $M_\odot$/yr, and J0910+0218 spans 50-397 $M_\odot$/yr, and J0907+0423 is constrained at the 95\% level to have SFR$<170$ $M_\odot$/yr (see Table \ref{tab:halpha}). 

These wide constraints are even broader than the systematic uncertainties from the UV-to-IR SED spectrophotometric fitting presented in \cite{Setton2025_squiggle}, making no progress in resolving the dust-star geometry. Additionally, as previously discussed, these measurements are subject to significant unaccounted for uncertainty due to the systematic effect of the aperture correction, which assumes that H$\alpha$ and H$\beta$ emission scale linearly with the continuum, which could result in over-estimated line fluxes. However, these measurements also are systematically uncertain because they only probe regions of the ISM where $\tau_{\mathrm{H}\alpha}\lesssim1$, which could result in an under-estimate of the total line flux by preferentially missing the most obscured regions. As such, we consider these measurements to be highly unreliable, motivating a resolved push into the near-IR. We proceed forward instead utilizing NIRSpec spectroscopy, which has the significant advantage of increased sensitivity, requires no aperture correction, and accesses a suite of hydrogen recombination lines which are less sensitive to dust attenuation, to determine the true star formation rates of these galaxies.

\subsection{Spatially resolved NIR line fitting} \label{subsec:NIRSpec_fitting}

\input{tables/nir_lines_table}

\begin{figure*}
    \centering
    \includegraphics[width=\textwidth]{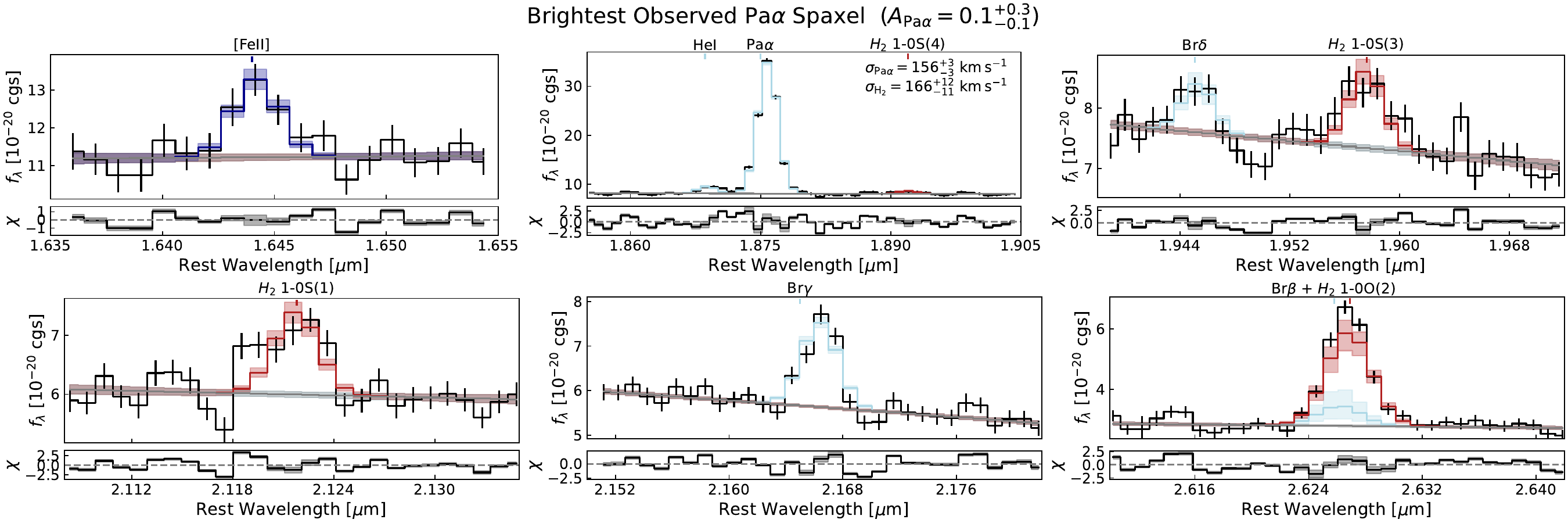}
\includegraphics[width=\textwidth]{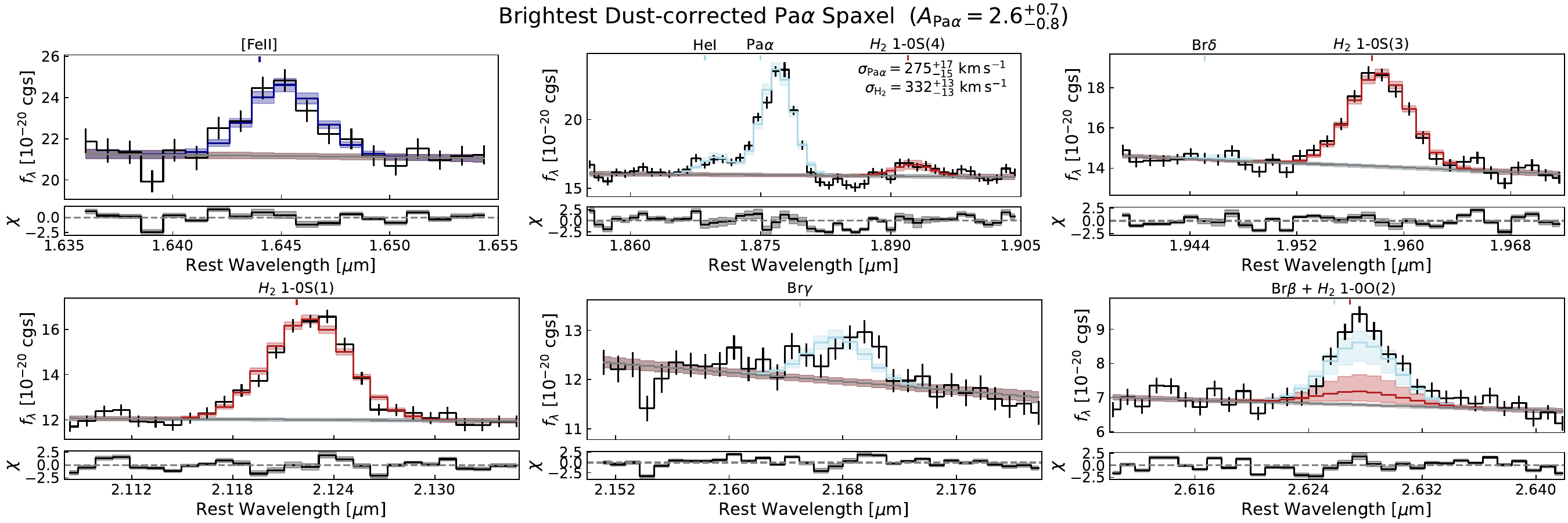}
    \caption{A demonstration of our line fitting procedure for J1157+0132, highlighting some (but not all) of the lines we fit in each individual $\sim$700 pc spaxel. On top, we show the fit to the spaxel with the brightest observed Pa$\alpha$ luminosity, which is consistent with no attenuation and low ($\sim150$ km/s) dispersion. Below, we show the fit to the spaxel with the brightest \textit{dust-corrected} Pa$\alpha$ luminosity, with $A_{\mathrm{Pa}\alpha}=2.6\pm_{0.8}^{0.7}$ and high ($\sim300$ km/s) dispersion. Note that while the fits to Br$\beta$ and $H_2$ 1-0O(2) are formally independent and tied to the kinematics of the HI and $H_2$ lines, respectively, we do not attempt to decompose these lines and only the total line flux of that complex enters into our analysis.}
    \label{fig:line_fitting}
\end{figure*}

The near-IR spectra of \squiggle post-starburst galaxies are populated with a wide range of hydrogen recombination and ro-vibration $H_2$ emission lines, often quite close to one another in wavelength (see Figures \ref{fig:J1157_fulldata}, \ref{fig:J0910_fulldata}, and \ref{fig:J0907_fulldata}). In order to obtain line maps for the recombination lines that are sensitive to dust attenuation, we jointly model the emission in individual spaxels by fitting for a wide suite of these features in small chunks of wavelength, modeling the continuum with a simple linear function. To do so, we utilize the \texttt{unite} package \citep{Hviding2025_LRD}, which uses a Hamiltonian Monte Carlo approach to rapidly explore the posteriors of high-dimensional fits and also allows for priors to easily be placed on entire groups of lines. In every spaxel, we fit the following families of lines, tying their redshifts and FWHM but leaving their normalization free: (1) Hydrogen and Helium recombination lines, (2) Molecular hydrogen lines. Lines in a given category have grouped kinematics, both in line center and in velocity dispersion. Additionally, we fit for [Fe\,\textsc{ii}] 1.6440$\mu$m, which we group kinematically with the Hydrogen and Helium recombination lines. The full list of lines we fit is listed in Table \ref{tab:nir_lines}.

This line list is by no means exhaustive. For example, there is a clear detection in all three galaxies of a suite of $H_2$ Q-series transitions at rest-frame $\sim2.4$ $\mu$m that we are not accounting for in our fits. However, these lines occupy a much more complicated part of the spectrum that is also populated by strong CO molecular bandheads in absorption, tracing the contribution to the continuum from evolved stars. As such, we defer the interpretation of the full hydrogen series to a future work, and instead fit lines only when they are fairly isolated or are essential due to their proximity to Hydrogen and Helium recombination lines.

We fit all spaxels with a median continuum signal-to-noise-per-wavelength-element greater than 2. In order to easily allow the posteriors for line fluxes to define detections, we set a prior that allows for negative line fluxes for almost all lines. One notable exception to this choice is the blend of Br$\beta$ and $H_2$ 1-0 O(2), which are offset by $\sim100$ km/s. Because this offset is similar to the spectral resolution of NIRSpec with the G395M grating, we do not attempt to deblend these features. Instead, we leave both lines free, with velocity and dispersion values fixed to the values for the rest of their species, but with the amplitude restricted to be positive. In practice, this allows us to fit for the total line flux of this blend. Future work will attempt to leverage information from the ro-vibrational $H_2$ spectral line energy distributions to deblend these features by placing stronger priors on the observed $H_2$ 1-0 O(2) line flux, but in this work, only the total line flux of the feature enters into our analysis.

In Figure \ref{fig:line_fitting}, we showcase our fitting to two spaxels from J1157+0132: the spaxel with the highest observed Pa$\alpha$ flux (top) and the spaxel with the lowest Pa$\alpha$/Br$\gamma$ ratio where both lines are detected at the 3$\sigma$ level (bottom). These spaxels highlight the diversity in line ratios on the $\sim700$ parsec spatial scales probed by our data, where spaxels with strong hydrogen recombination emission and Pa$\alpha$/Br$\gamma$ in line with Case B exhibit narrow recombination lines and fairly weak ro-vibrational $H_2$ emission, whereas spaxels with Pa$\alpha$/Br$\gamma$ well below Case B tend to have much higher dispersion and much stronger ro-vibrational $H_2$ emission. 

\subsection{Spatially mapping the dust attenuation}

\begin{figure*}
    \centering
\includegraphics[width=\textwidth]{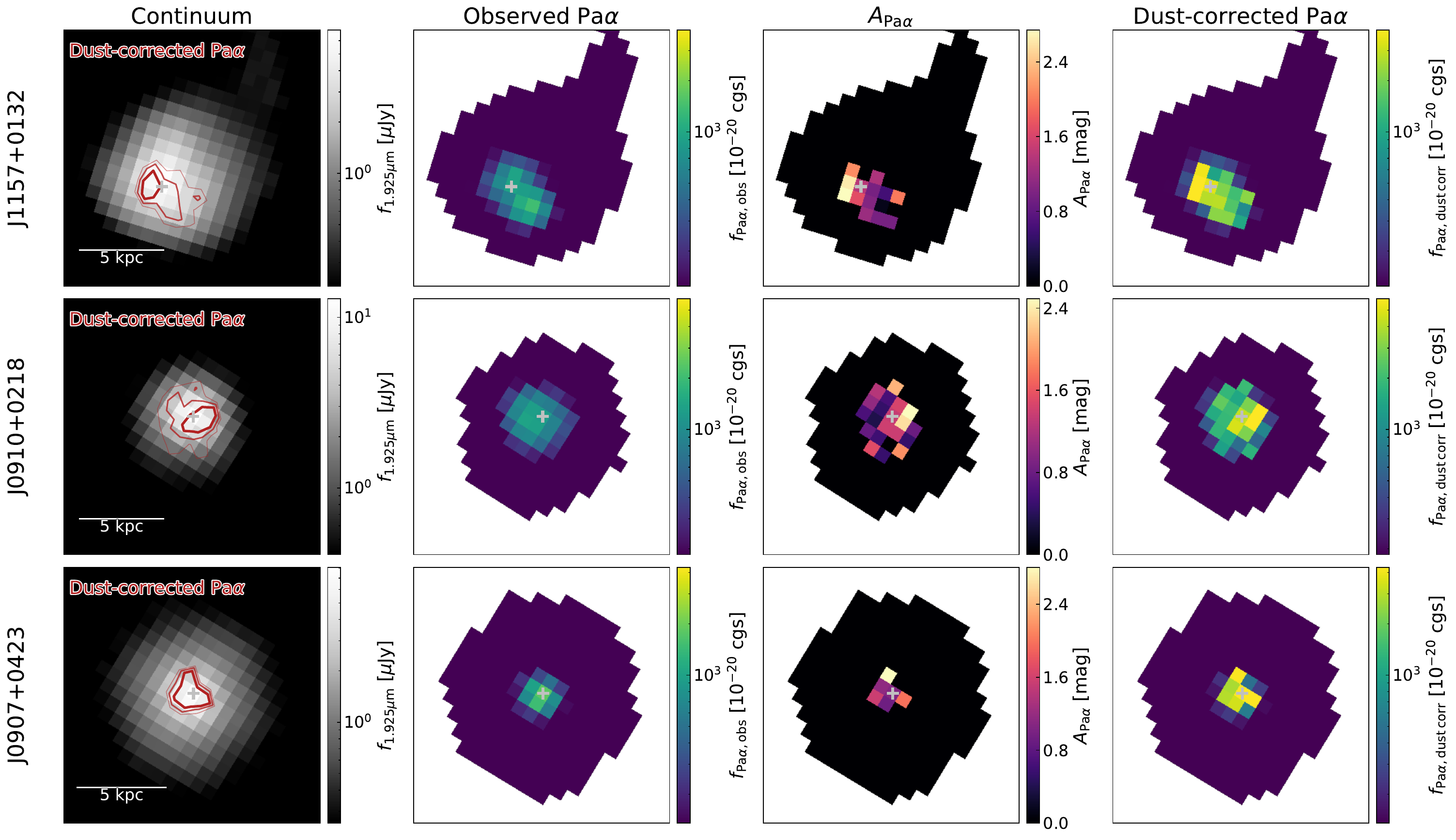}
    \caption{(First column): The collapsed 1.925 $\mu$m flux maps of the three galaxies in this work, with overlaid contours showing the dust-corrected Pa$\alpha$ flux (red), bounding the 10, 25, and 50\% levels. (Second column): The observed Pa$\alpha$ flux. (Third column): The median Pa$\alpha$ attenuation maps used to correct the observed Pa$\alpha$ flux. (Fourth column): The dust-corrected Pa$\alpha$ flux, corrected for attenuation (the same as the contours shown in the first column). In each panel, we mark the brightest continuum spaxel with a +. In all cases, the dust-corrected Pa$\alpha$ luminosity is heavily peaked in the most attenuated regions, which are near the galaxy centers.} 
    \label{fig:dust_maps}
\end{figure*}

Having successfully obtained posteriors for the line fluxes in all spaxels, we now turn to measuring the attenuation of Pa$\alpha$ to obtain the dust-corrected Pa$\alpha$ luminosity that will serve as our tracer of the instantaneous star formation rate. In order to correct the emission for underlying absorption, we adopt the models from \cite{Setton2025_squiggle}, reconstructed with the C3K libraries \citep{Conroy2012_C3K} to utilize its NIR resolution. The models predict EW(Pa$\alpha$)=[$1.21\pm^{0.14}_{0.18}$, $1.05\pm^{0.12}_{0.18}$, $1.45\pm^{0.19}_{0.20}$] $\mathrm{\AA}$ and EW(Br$\gamma$)=[$0.99\pm^{0.04}_{0.05}$, $1.06\pm^{0.06}_{0.05}$, $0.94\pm^{0.07}_{0.06}$] $\mathrm{\AA}$ for J1157+0132, J0910+0218, and J0907+0423, respectively. We adopt the median value for this absorption, and fix EW(Br$\beta$)=EW(Br$\gamma$), as even the C3K models lack the resolution to predict Br$\beta$ equivalent width. In practice, this choice matters very little, as Pa$\alpha$ and the Br$\beta$+$H_2$ 1-0 O(2) are typically high enough in equivalent width that the corrections are minor, and only Br$\gamma$ is highly influenced by the absorption correction. However, all these inferences are highly dependent on the exact prescription for thermally pulsing AGB stars in models \citep[e.g.,][]{Lu2025_AGB}, and therefore our predicted absorption line equivalent widths may be systematically uncertain. Additionally, the assumption that the absorption line equivalent widths do not vary spatially across the galaxy may be incorrect, though rest-optical spectroscopy of \squiggle post-starburst does suggest that color gradients on $\sim$kpc-scales are fairly flat in these galaxies \citep{Hunt2018, Setton2020}. Future detailed full spectral fitting of these spectra to understand the exact composition of starlight that dominates as a function of wavelength will better constrain these absorption profiles in individual spaxels.

With absorption corrected line fluxes, we have what we need to measure the deviations in the recombination lines from Case B \citep[intrinsic Pa$\alpha$/Br$\gamma=12.07$, Pa$\alpha$/Br$\beta=7.43$,][]{Hummer1987}, and therefore to measure the dust attenuation of Pa$\alpha$. The limiting factor in conducting this analysis is the signal-to-noise in the Br$\gamma$ emission, the weakest included line. As such, we only utilize spaxels where both Br$\gamma$ and Pa$\alpha$ are detected at the 3$\sigma$ level. As an additional robustness check, we utilize the combined Br$\beta$ and $H_2$ 1-0 O(2) fluxes measured in the previous sections. We take Br$\beta_\mathrm{max}$ to be equal to the total flux of that complex, including all contributions from Br$\beta$ and $H_2$ 1-0 O(2). We assume no dust correction for any draw from the posterior for our line fluxes where the ratio of Pa$\alpha$/Br$\beta_\mathrm{max}$ is consistent with being in excess of Case B assumptions, even when Pa$\alpha$/Br$\gamma$ is below Case B, safeguarding us against spurious Br$\gamma$ detections.

After performing those cuts, we measure the posterior in $A_{\mathrm{Pa}\alpha}$ in all remaining spaxels. This measurement is subject to the assumed slope of the dust curve between Pa$\alpha$ and Br$\gamma$. \cite{Cardelli1989} parametrizes the law at these wavelengths as $A(\lambda)\propto \lambda^{- \alpha}$, with $\alpha=1.61$. However, more recent studies have favored a steeper IR extinction law, with $\alpha\sim2.1$ \citep[e.g.,][]{Fritz2011_nearIR_dust, Wang2019_dust_law}, with studies toward the galactic center suggesting even more extreme values \citep[$\alpha\sim2.23$, e.g.,][]{Nogueras-Lara2020_NIR_extinction}. Furthermore, unresolved studies of high-z star forming galaxies also favor steeper-than-Cardelli near-IR curves \citep[though not quite at wavelengths as red as Pa$\alpha$ and Br$\gamma$, see][]{Reddy2026_dustlaw}. Therefore, here we adopt a power law-parameterization of the near-IR dust law with $\alpha=2.11$ \citep{Fritz2011_nearIR_dust}, which was measured using emission lines toward the galactic center, noting that the systematic uncertainty in $\alpha$ ranging from $\sim 1.6-2.2$ imparts a factor of $\sim1.5$ total uncertainty on $A_{\mathrm{Pa}\alpha}$, and a factor of $\sim2$ in the star formation rate.

As shown in Figure \ref{fig:line_fitting} using J1157+0132 as an illustrative example, there is a wide range in Pa$\alpha$ attenuation throughout an individual galaxy, ranging from essentially none (top) to $A_{\mathrm{Pa}\alpha}$ as high as 2 (bottom). Summing over all pixels and comparing the observed and dust-corrected fluxes, we find that the effective $A_{\mathrm{Pa}\alpha}$ is $1.15\pm^{0.18}_{0.16}$, $1.09\pm^{0.11}_{0.12}$, and $1.04\pm^{0.20}_{0.17}$ for J1157+0132, J0910+0218, and J0907+0423, implying a factor of 2-3 correction to the observed Pa$\alpha$ is required to obtain the dust-corrected flux. This implies that Pa$\alpha$ is not a dust-free tracer of star formation, as is commonly assumed \citep[see also][which reports a similarly high $A_{\mathrm{Pa}\alpha}$ in a systematic study of Pa$\alpha$/Br$\beta$/Br$\alpha$ in a general sample of IR luminous galaxies]{Vanicek2026_PaA_BrA}.

Under the assumption of a \cite{Wang2019_dust_law} dust law, which stitches together the steeper-than-Cardelli near-IR dust law favored by recent observations with the optical form of the \cite{Cardelli1989} dust-law ($A_V = 9.87 A_{\mathrm{Pa}\alpha}$), our $A_{\mathrm{Pa}\alpha}$ measurements imply $A_{V,\mathrm{nebular}} = 11.3\pm^{1.8}_{1.6}, 10.8\pm^{1.1}_{1.2}, \ \mathrm{and} \ 10.3\pm^{2.0}_{1.7}$ for J1157+0132, J0910+0218, and J0907+0423, respectively. While highly uncertain (and subject to UV-to-IR functional form of the assumed dust law), these values are higher than the effective nebular $A_{V}$ implied by the H$\alpha$/H$\beta$ Balmer decrement and the \cite{Setton2025_squiggle} SED fits. However, this inconsistency is expected given the finding that the dustiest spaxels have $A_{\mathrm{Pa}\alpha}\sim2$. Those regions of the galaxies are optically thick to H$\alpha$, and are completely missed in both the integrated $H\alpha$ and $H\beta$ measurements that are heavily weighted toward the least dusty regions. It is likely that under such extreme obscuration, a two screen model with a single dust law fails to capture the full behavior across such wide wavelength, and a model which accounts for the covering fraction of highly obscuring dust \citep[e.g.,][]{Reddy2026_dustlaw} is more applicable to this data. Nevertheless, if a \cite{Fritz2011_nearIR_dust} curve is applicable, our $A_{\mathrm{Pa}\alpha}$ measurements should provide robust dust-corrections for Pa$\alpha$ and nearby lines.

In Figure \ref{fig:dust_maps}, we visualize this correction by showing the two dimension continuum maps (first column), observed Pa$\alpha$ maps (second column), $A_{\mathrm{Pa}\alpha}$ maps (third column), and dust-corrected Pa$\alpha$ maps (fourth column). In all three galaxies, the dust-corrected Pa$\alpha$ luminosity is concentrated in the highly attenuated galaxy cores, which are compact compared to the 1.925 $\mu$m continuum. While we defer a quantitative structural analysis accounting for the NIRSpec IFU PSF to a future work, qualitatively, it appears that all three of these galaxies are host to centrally concentrated Pa$\alpha$ emission behind optically thick dust on $\sim1$ kpc scales.

\subsection{Accounting for non-star forming ionization} \label{subsec:non-sf-ionization}

Before using the dust-corrected Pa$\alpha$ luminosity to measure the star formation rates, we first turn our focus to the possibility of other ionization sources that could contribute to Pa$\alpha$ and inflate inferred star formation rates. Specifically, given the massive, post-merger nature of these galaxies, we are concerned about the impact of AGN activity and shocks. This concern is motivated by rest-optical observations of post-starburst galaxies. A significant fraction of the youngest \squiggle galaxies (including J1157+0132) have [OIII]/H$\beta$ ratios consistent with AGN in the mass-excitation diagram \citep[$\sim20\%$ of galaxies in the lowest $\mathrm{D_{n4000}}$ quantile, see][]{Greene2020}, and the two sources studied in this work with detected [NII] and H$\alpha$ have high ratios that indicate the presence of AGN or LINER emission \citep[consistent with the broader \squiggle sample, see][]{Zhu2025}. Additionally, all three of the sources have strong [OII] emission, a potential indicator of shocks \citep[e.g.,][]{Alatalo2016b, Alatalo2016a}. As such, it would be naive to interpret the Pa$\alpha$ luminosity as purely driven by star formation.

The ratio of [Fe II] emission to hydrogen recombination emission is sensitive to the dominant source of ionization in a galaxy, with lower values seen in galaxies where star formation dominates and higher values seen in AGN and LINERs \citep{Larkin1998_NIR_BPT, Dale2004_NIR, RodriguezArdila2004_FeII_H2, RodriguezArdila2005_FeII_H2, Riffel2013}. [FeII]1.674$\mu$m is covered in two of our galaxies (J1157+0132 and J0910+0218); as such, we focus on those sources. To estimate the contribution from non-star formation driven ionization in each spatial pixel, we assume that the observed line fluxes are a linear combination of flux from star formation- and AGN/LINER/shock-dominated regions, each with a fixed intrinsic ratio of [Fe II]-to-hydrogen recombination emission. Following this assumption, we estimate $f_\mathrm{non-SF}$, the fraction of the Pa$\alpha$ that is produced from non-star formation driven ionization, using the following formulation:

\begin{equation}
    f_\mathrm{non-SF} = \frac{([\mathrm{FeII}]/\mathrm{Pa}\alpha)_\mathrm{obs} - ([\mathrm{FeII}]/\mathrm{Pa}\alpha)_\mathrm{SF}}{([\mathrm{FeII}]/\mathrm{Pa}\alpha)_\mathrm{non-SF} - ([\mathrm{FeII}]/\mathrm{Pa}\alpha)_\mathrm{SF}}
\end{equation}

We define the constant contribution from star forming and non-star forming ionization empirically for each galaxy. For $([\mathrm{FeII}]/\mathrm{Pa}\alpha)_\mathrm{SF}$, we take the lowest well-detected [FeII]/Pa$\alpha$ spaxel ($\sim0.05$) to be representative of pure star formation. This is consistent with expectations from the literature, as \cite{Riffel2013} proposes that [Fe II] 1.257 $\mu$m/Pa$\beta < 0.6$ delineates gas dominated by star formation. Under Case B conditions for hydrogen (Pa$\alpha$/Pa$\beta=2.05$) and using the theoretical ratio of [Fe II] 1.644 $\mu$m/[Fe II] 1.257 $\mu$m $=0.74$ \citep{Nussbaumer1988_FeII}, this boundary corresponds to [Fe II] 1.644/Pa$\alpha < 0.2$. For $([\mathrm{FeII}]/\mathrm{Pa}\alpha)_\mathrm{non-SF}$, we take the highest [FeII]/Pa$\alpha$ spaxel (0.82 in J1157+0132 and 0.46 in J0910+0218) to be a general representation of non-star forming ionization. As a test of our ratio for this assumption, we utilize the 3MdB database shock models \citep{Alarie2019_shocks}, which are computed using the \cite{Allen2008_mappings} parameters. We find that shock models span a wide range in [FeII]/Pa$\alpha$ (0.5–7 depending on velocity and magnetic field strength). As such, our assumed ``pure" non-star forming ionization value is conservative, as a higher assumed $([\mathrm{FeII}]/\mathrm{Pa}\alpha)_\mathrm{non-SF}$ would result in lower $f_\mathrm{non-SF}$ and a larger contribution to the total Pa$\alpha$ luminosity from pure star formation. 

For each individual spaxel, we evaluate $f_\mathrm{non-SF, max}$ using the formulation above. We then correct all spaxels for this value, and evaluate the galaxy-wide effective $f_\mathrm{non-SF,max}$ by summing those pixels and comparing to the uncorrected flux. We find $f_\mathrm{non-SF,max}$ to be 42\% and 32\% in J1157+0132 and J0910+0218, respectively, suggesting that a significant fraction of the total Pa$\alpha$ luminosity could result from non-star forming sources. Given that we do not target [FeII] in J0907+0423, we take $f_\mathrm{non-SF,max}$ to be 37\%, the median of our measurements of the other two galaxies. Throughout this work, we carry these maximal corrections along to bracket the full range in possible star-formation driven ionization of the observed Pa$\alpha$.

\subsection{The global star formation rate} \label{subsec:sfr}

\input{tables/paa_sfr_table}

We now turn to measuring the total Pa$\alpha$ luminosity and the star formation rate. We sum all Pa$\alpha$ flux measurements, after dust corrections, to obtain a posterior for the dust-corrected Pa$\alpha$ luminosity, which we also correct with the $\sim35\%$ $f_\mathrm{non-SF,max}$ corrections discussed in the previous section. Following \cite{Neufeld2024_PaA} (and references therein), we measure instantaneous star formation rates using the following calibration for a \cite{Kroupa2001} IMF (again noting the equivalence for the \cite{Chabrier2003} to \cite{Kroupa2001} conversion):

\begin{equation}
    \mathrm{SFR} \ [M_\odot / \mathrm{yr}] = 4.6 \times 10^{-41} \times L_{\mathrm{Pa}\alpha} \ [\mathrm{erg/s}]
\end{equation}

These results, along with associated uncertainties, are tabulated in Table \ref{tab:paa_sfr}, for both star formation dominated Pa$\alpha$ and maximal non-star formation driven ionization of Pa$\alpha$.

\begin{figure*}
    \centering
    \includegraphics[width=\textwidth]{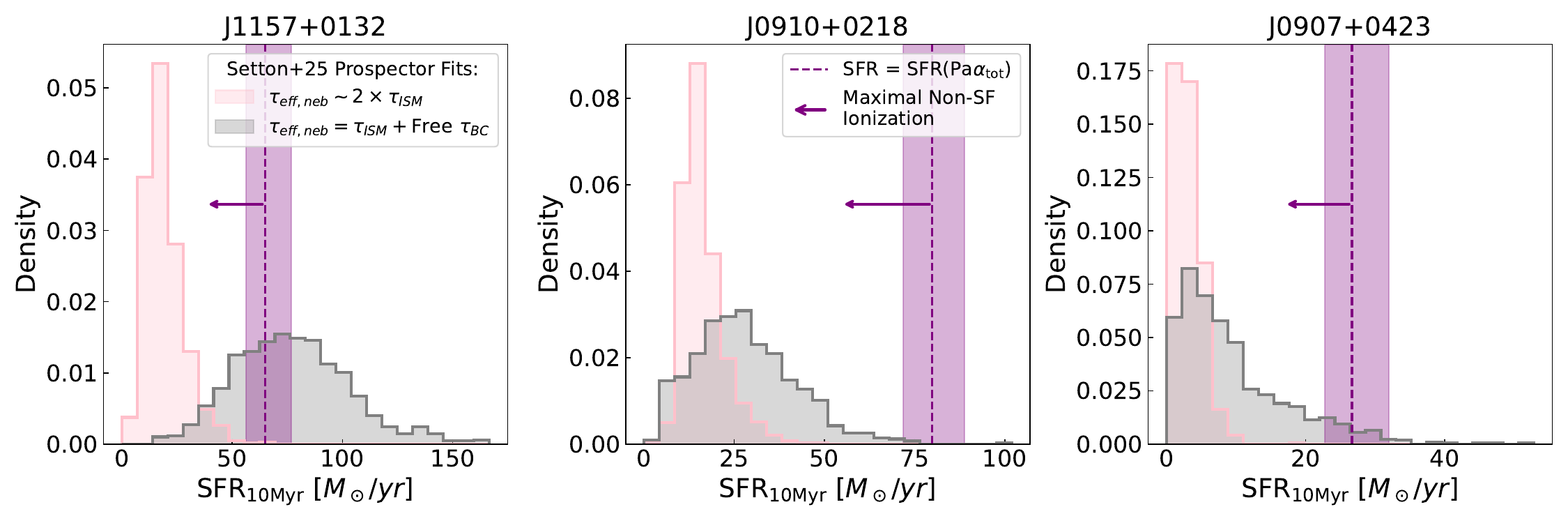}
    \caption{The Pa$\alpha$-derived star formation rates measured in this work (purple, with an arrow indicating the maximal correction to the star formation rate in the scenario where $\sim30-40\%$ of the ionization is due to non-star formation sources), plotted alongside the posteriors for the 10 Myr star formation rates from the UV-to-IR spectrophotometric fits presented in \cite{Setton2025_squiggle}. The posteriors for the $\tau_{BC}$ tied fits that placed a strong prior on the birth cloud dust attenuation being closely tied to be $2\times$ the diffuse ISM attenuation are shown in pink, and the $\tau_{BC}$ free fits in which birth cloud dust was decoupled from the ISM dust are shown in grey. The instantaneous star formation rate measured from the $\tau_{BC}$ tied fits is a factor of $2-3\times$ lower than the Pa$\alpha$-derived star formation rates, whereas the (quite uncertain) $\tau_{BC}$ free fits encompass the Pa$\alpha$ star formation rates, especially under the assumption of a maximal correction due to non-star formation driven ionization.}
    \label{fig:SFR_comp}
\end{figure*}

We now compare these instantaneous star formation rates to those obtained from UV-to-IR SED fitting. In \cite{Setton2025_squiggle}, we fit two sets of models to each post-starburst. In both models, we assumed a \cite{Charlot2000} two-screen dust model, where all stellar populations see dust with V-band optical depth $\tau_{ISM}$ \citep[with an attenuation law set with the \cite{Noll2009} power-law modification of a][dust law]{Calzetti1997}, and where stars with $t<10$ Myr see an additional dust screen with V-band optical depth $\tau_{BC}$. One model imposed a strong prior that is common in studies of massive, (presumably) quenched galaxies, where $\tau_{ISM}\sim\tau_{BC}$, a relation that is broadly found to be true in main sequence massive star forming galaxies \citep[e.g.,][]{Price2014}. The other model allowed significantly more birth cloud attenuation by allowing for a free $\tau_{BC}$ that is decoupled from $\tau_{ISM}$. In both fits, the total attenuated luminosity was re-emitted through dust emission parameterized with \cite{Draine2007} models, primarily constrained by WISE W3/W4 and ancillary Herschel observations. Throughout this work, we use these fits as our benchmarks for the measurement of the star formation rate with Pa$\alpha$, and refer to the fits as $\tau_{BC}$ tied and $\tau_{BC}$ free, respectively.

In Figure \ref{fig:SFR_comp}, we show our new measurements as purple lines (with an arrow showing magnitude of an additional $\sim30-40\%$ correction for the maximum non-star forming contribution to Pa$\alpha$), overlaid on the posteriors for the 10 Myr star formation rates from the \cite{Setton2025_squiggle} fits, with the $\tau_{BC}$ tied fits shown in pink and the $\tau_{BC}$ free fits shown in grey. It is immediately clear that the flexible fits that allow for greater effective nebular attenuation are necessary to capture the high star formation rates we measure here. In fact, even these fits, which still placed a prior that disfavored $A_{V,\mathrm{total}} \gtrsim 5$ may be insufficient if our correction to Pa$\alpha$ for the effects of shocks or AGN activity is too strong. In the case of no correction for non-star formation driven ionization, J0910+0218 and J0907+0423 are only barely consistent with this posteriors, which have an upper limit in the star formation rate that is set by energy balance of the attenuated luminosity with available IR limits \citep[see][]{Setton2025_squiggle}. However, with the correction for non-star formation driven ionization described in Section \ref{subsec:non-sf-ionization}, the star formation rates fall much more comfortably in the posterior of the $\tau_{BC}$ free fits, and are still a factor of 2-3$\times$ higher than the $\tau_{BC}$ tied fits, as is also the case in J1157+0132. As such, it is clear that for the most CO-luminous \squiggle post-starburst galaxies, fits that did not account for the possibility of such high dust attenuation in birth clouds \citep[including our own fits and H$\alpha$ analysis in][]{Suess2022a, Bezanson2022a, Zhu2025} systematically under-estimated the true star formation rates of the most CO luminous systems by as much as a factor of 5. While it is not clear that accounting for such birth clouds is necessary for the more general population of CO-undetected post-starburst galaxies (or even the more moderately CO-luminous \squiggle galaxies), we conclude the most CO-luminous post-starburst galaxies still host significant star formation that is obscured in the rest-optical by strong dust attenuation.

We acknowledge the possibility that our Pa$\alpha$/Br$\gamma$ dust corrections are susceptible to a failure mode similar to that of the H$\alpha$/H$\beta$ Balmer decrements, where we are only penetrating to the surface where dust is optically thick at 2 $\mu$m. Thus, formally, our Pa$\alpha$ star formation rates are lower limits. However, based on extrapolations of the WISE (and, in the case of J1157+0132 and J0910+0218 Herschel/SPIRE 250 $\mu$m) photometry with \cite{Draine2007} dust models, the \cite{Setton2025_squiggle} fits measure total IR luminosities that can be used to place a ceiling on the $\sim100$ Myr averaged star formation rate. Adopting the \cite{Murphy2011} calibration for the total IR luminosity (SFR [$M_\odot$/yr] = $3.88\times10^{-44}$ $L_\mathrm{IR} \ \mathrm{[erg/s]}$), we find $95\%$ upper limits of 196, 134, and 61 $M_\odot$/yr for J1157+0132, J0910+0218, and J0907+0423, respectively, leaving some room for true star formation rates that are a factor of $\sim 3–5/ 2–3/2–4 \times$ higher than the Pa$\alpha$-inferred rates, similar to a trend seen in ULIRGs at similar redshift \citep{Vanicek2026_PaA_BrA}. However, this comparison is extremely conservative due to the aforementioned issue of different timescales, where Pa$\alpha$ traces $\sim$10 Myr timescales, but $\sim100$ Myr timescales are relevant for $L_\mathrm{IR}$-derived tracers. In the following section, we will show that the \cite{Setton2025_squiggle} fits suggest sustained peak star formation rates that are in excess of the IR luminosity-derived limits by a factor of $\sim2$, suggesting that we are indeed operating in the situation described in \cite{Leja2019} where a significant contribution to $L_\mathrm{IR}$ from older stellar populations biases inferences of the instantaneous star formation rate under the assumption of a constant star formation on a $<100$ Myr timescale. As such, we proceed under the assumption that our star formation rates are secure, but note that future observations with a facility like MIRI/MRS (probing truly dust-insensitive features like Hu$\alpha$ and [Ne II] 12.81 $\mu$m) could potentially allow us to conclusively rule on whether the Pa$\alpha$-derived star formation rates are underestimates of the true instantaneous star formation rate.

\section{Placing CO-luminous post-starburst galaxies in an evolutionary context} \label{sec:discussion}

Armed with secure star formation rate measurements for our post-starburst galaxies, and the knowledge that the star formation histories from \cite{Setton2025_squiggle} encompass solutions that can simultaneously match the observed star formation rates in the most CO-luminous systems while also fitting the full UV-to-IR SED and rest-optical spectroscopy, we now turn to placing these galaxies in an evolutionary context. First, we revisit the trend in the CO luminosity and the age of the recent starburst that was quantified in \cite{Bezanson2022a} and \cite{Setton2025_squiggle} in the context of these new measurements. Then, we will propose that the CO-luminous phase of these post-starburst galaxies represents the last stages of their major starburst, suggesting an evolutionary sequence from starburst to post-starburst that can be explored with complete massive galaxy samples that span this entire sequence.

\subsection{Can the rapid depletion of molecular gas reservoirs be explained by buried star formation alone?} \label{subsec:tdep}

First, we seek to interpret the luminous CO in these galaxies in the context of these new star formation rates. The CO luminosity (or $M_{H_2}$, under the assumption of a universal $\alpha_{CO}$) in \squiggle correlates strongly with the age of the most recent starburst. The systems which are not detected in CO(2-1) have characteristic times since quenching ($t_{q}$) of $\sim300$ Myr \citep{Bezanson2022a, Setton2025_squiggle}. This suggests that if star formation were not present in the young CO luminous galaxies we observe here, they would need a similar amount of time to present as the older CO non-luminous system, implying rapid depletion. However, in this work, we demonstrate even under the conservative assumption that non-star formation-driven ionization contributes $\sim1/3$ of the total Pa$\alpha$ luminosity that these systems are not passive. As such, without knowing exactly how long their present period of star formation will be sustained, it is difficult to directly connect these galaxies to their quenched descendants. Nevertheless, these new, more secure star formation rates allow us to re-visit the depletion times of these systems and to ask how long they could live in this CO luminous phase before passively evolving into truly quenched quiescent systems, under a range of assumptions.

We begin by making the standard assumption from previous works: a Milky Way CO-to-H$_2$ conversion and an assumption of thermalized CO that have been adopted in past studies of these samples and other co-eval quiescent galaxy samples \citep[$\alpha_\mathrm{CO}=4$ and $r_{21}=1$, e.g.,][]{Suess2017, Spilker2018, Bezanson2022a, Woodrum2022, Setton2025_squiggle}. In this context, the depletion times ($t_\mathrm{dep} = M_\mathrm{H_2}/\mathrm{SFR}$) that we measure--in this case, conservatively neglecting the potential for $\sim35\%$ non-star-formation-driven ionization--are $1.7\pm_{0.3}^{0.3}$, $0.8\pm_{0.1}^{0.1}$, and $2.2\pm_{0.4}^{0.4}$ Gyr for J1157+0132, J0910+0218, and J0907+0423 respectively. This implies that these galaxies could spend as much as 1-2 Gyr burning through their remaining fuel at these observed star formation rates before quenching (and, crucially, not rapidly), making them incompatible with direct progenitor status of the CO-luminous systems. 

Nevertheless, we hold that it is still a distinct possibility that there is a direct evolutionary link on the observed timescale of $\sim$a few hundred Myr between the CO luminous and CO undetected post-starburst galaxies. First, astration is likely not the only mechanism at play in the youngest \squiggle galaxies. Luo et al. 2026 submitted shows that the youngest \squiggle\ galaxies, which preferentially are CO luminous, host neutral gas outflows with mass outflow rates $\sim40$ $M_\odot$/yr, comparable to the star formation rates seen in systems we measure here. The present day star formation rates are not high enough to explain these $\sim200-400$ km/s outflows, and they are likely relics launched during the peak of the starburst, where AGN activity may also have peaked. Such outflows are indeed seen in compact starbursts which are likely progenitors of the \squiggle galaxies \citep[][see also the following section]{Tremonti2007,Sell2014,Geach2014,Geach2018,Perrotta2023_outflow, Perrotta2024}, making the removal of significant gas reservoirs through such a mechanism appealing. Accounting for these outflows could lower the depletion times by a factor of $\sim2$.

\begin{figure*}
    \centering
    \includegraphics[width=\textwidth]{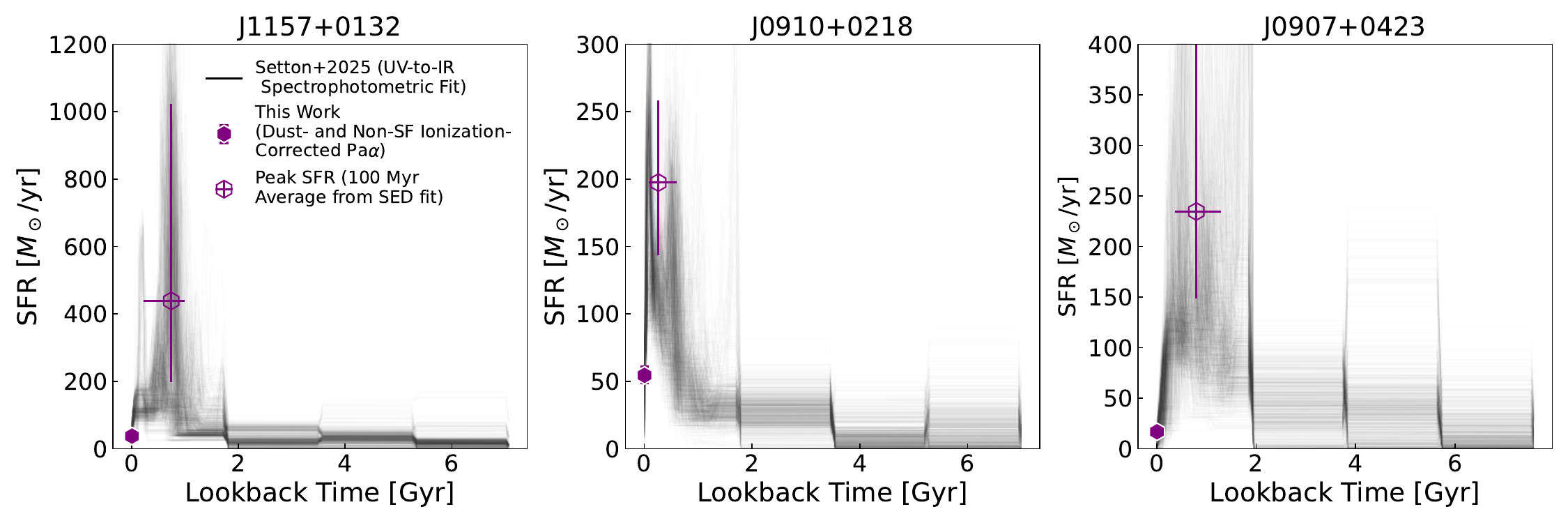}
    \caption{Individual draws for the star formation histories \citep[as measured in the $\tau_{BC}$ free fits from][]{Setton2025_squiggle} for each of the galaxies presented in this work, smoothed with a 100 Myr kernel. Overlaid are the Pa$\alpha$-derived star formation rates (filled purple hexagon, corrected for both dust and maximal non-star-formation-driven ionization) and the peak star formation rate from those fits (open purple hexagon) for each of the galaxies. All three galaxies, while presently star forming, are consistent with a significant recent drop (0.5-1 dex) in their star formation rate within the last few hundred Myr.}
    \label{fig:SFHs}
\end{figure*}

The other distinct factor in the depletion time we highlight is the molecular gas mass. The central dusty spaxels, which host the most abundant star formation, are also host to the strongest ro-vibrational Hydrogen transitions (see the bottom example in Figure \ref{fig:line_fitting}) with very high ($\sim1000$ km/s full-width half-maximum) dispersion and evidence for shocks in their elevated [Fe\,\textsc{ii}]/Pa$\alpha$ ratios (see Section \ref{subsec:non-sf-ionization}). The central reservoirs in these post-merger systems could be very disturbed, leading to the exact conditions where a departure from the Milky Way bound-molecular cloud conditions that lead to $\alpha_{CO}\sim 4$ could occur, making an assumed $\alpha_{CO}\sim 0.8$ more appropriate \citep[e.g.,][]{Bolatto2013}. Additionally, the elevated CO SLEDs observed in many of the most CO luminous \squiggle galaxies \citep[typical $L'_\mathrm{CO(5-4)}/L'_\mathrm{CO(2-1)}\sim0.2-0.4$,][]{Donofrio2026_CO_excitation, Zanella2026_CO_excitation} are potential evidence of such conditions. While speculative, an additional factor of $\sim5$ reduction in the depletion time in the most CO luminous systems could be driven by the extreme conditions of the mergers \citep{Narayanan2011}. Mergers are indeed over-represented in the youngest, most CO-luminous \squiggle post-starburst galaxies \citep{Spilker2022, Verrico2023, Donofrio2025}, and in massive post-starburst systems generally \citep{Sazonova2021, Maltby2026_PSB_structure}.

Tallying all these effects together, one can imagine a factor of $\sim10$ reduction in the total depletion times we measure, resulting in depletion times on the order of $\sim100-200$ Myr rather than 1-2 Gyr, a decidedly shorter timescale that is much more consistent with the direct evolution from these CO luminous \squiggle galaxies into the CO undetected population, on a timescale that \textit{could} be referred to as rapid. While it is clear that star formation effects cannot alone account for this evolution \citep[in agreement with the conclusion of][]{Setton2025_squiggle}, with star formation rates now securely measured, pinning down both the $M_{H_2}$ and the mechanical sources of $\dot{M}_{H_2}$ is a crucial next step in better characterizing the depletion times--and therefore the future evolution--of these systems.

\subsection{Running back the clock: \squiggle at the peak of their starburst} \label{subsec:starburst}

\begin{figure*}
    \centering
    \includegraphics[width=\textwidth]{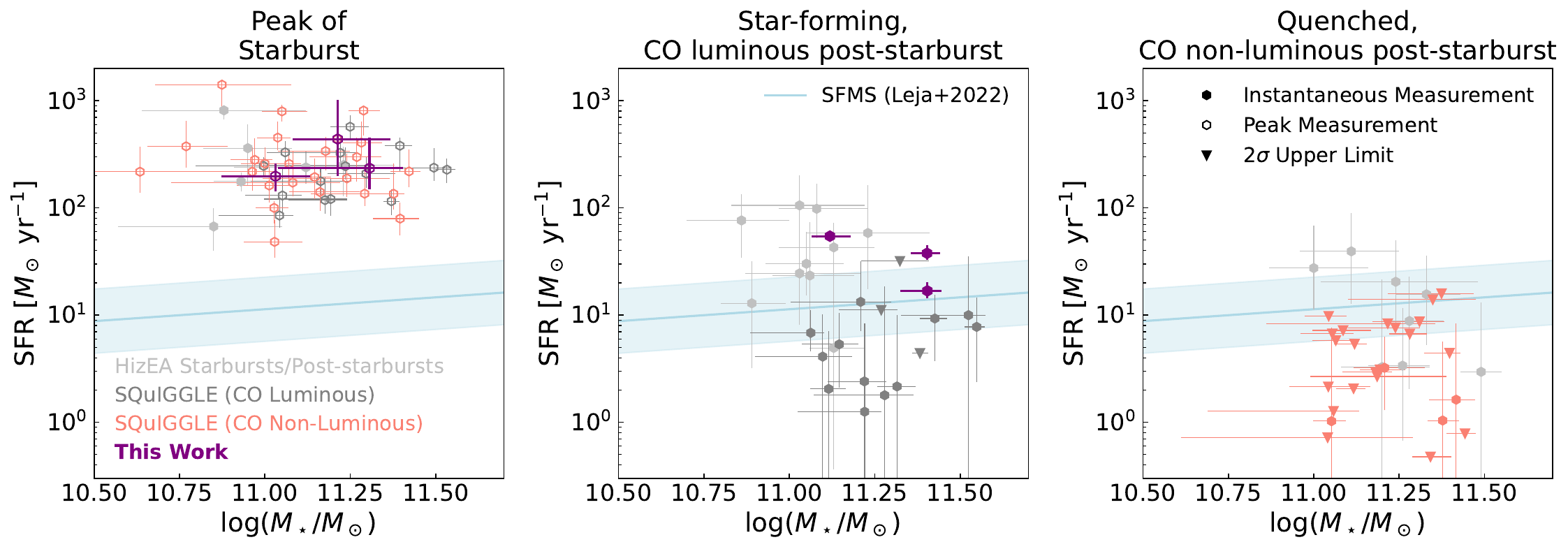}
    \caption{The star forming main sequence, showing the best fit relation from \cite{Leja2022}, with \squiggle galaxies shown at various stages of their evolution along a proposed sequence from starburst to initial, CO luminous decline to quiescence. In all panels, open hexes indicate a peak star formation rate measurement from the \cite{Setton2025_squiggle} star formation histories, while solid hexes indicate an ``instantaneous" measurement, either from those same fits or, in the case of the three galaxies studied in this work, from the dust- and non-star-formation-driven-ionization-corrected Pa$\alpha$ measurements. Instantaneous measurements are shown as a $2\sigma$ upper limit when the median of the posterior is $<1$ $M_\odot$/yr, the star formation rate reliability threshold established in \cite{Suess2022b}. CO luminous galaxies ($L'_\mathrm{CO} > 2.5 \times 10^{9}$ K km s$^{-1}$ pc$^{-2}$) are shown in grey (purple for measurements from this work), and CO non-luminous galaxies are shown in salmon. As a comparison sample, we also show the co-eval HIzEA starburst and post-starburst galaxies (Tremonti et al. in preparation, pink), which we divide between peak and decline based on their age ($t_\mathrm{light-weighted}<100$ Myr: left, $100<t_\mathrm{light-weighted}<200$ Myr: middle, $t_\mathrm{light-weighted}>200$ Myr: right). In the left panel, we show all galaxies at their peak star formation. In the center panel, we show the instantaneous measurements for the CO-luminous sample. In the right panel, we show the instantaneous star formation rates at the time of observation for the CO-non-luminous sample. We find that the full \squiggle sample is consistent with a similar magnitude of burst peak, with star formation rates of $\sim200$ $M_\odot$/yr, comparable to the instantaneous star formation rates of the HizEA compact starbursts. The young galaxies which we observe during the CO-luminous phase are in the final stages of their merger-driven bursts, with star formation rates that are in decline from their peak, spanning the main sequence. We speculate that the combined effect of this continued star formation and outflows, in conjunction with a non-Milky Way $\alpha_\mathrm{CO}$, may explain the direct evolution of these galaxies into the CO-non luminous, quiescent \squiggle galaxies.}
    \label{fig:SFMS}
\end{figure*}


We now turn our focus to the connection between the \squiggle post-starburst galaxies and their progenitors. Because the Pa$\alpha$ star formation rates in this work are contained in the posteriors of the $\tau_{BC}$ free \cite{Setton2025_squiggle} fits that allow for birth cloud dust in excess of what is expected from correlations between the ISM and birth cloud dust \citep[e.g.,][]{Price2014}, we adopt those fits as our models of choice for inferring the progenitor population of \squiggle galaxies. Drawing from the posterior of the fits, we calculate the star formation rate and stellar mass of the ``peak" of star formation by smoothing the star formation history with a 100 Myr kernel, finding the peak in the 2 Gyr before observation, and inferring the stellar mass and star formation rate at that peak. In Figure \ref{fig:SFHs}, we show this peak measurement on top of 1000 draws from the posterior for each of the three galaxies in this work, with the peak measurement and 1$\sigma$ uncertainties overlaid as an open hex and the Pa$\alpha$-derived instantaneous star formation rates shown as a filled hexagon. 

In all three galaxies, the appreciable measured star formation is still in significant decline from the peak of the bursts, which required sustained star formation for $\sim100$s of Myrs to produce the stellar population that exhibit the luminous, Balmer absorption dominated rest-optical spectra exhibited by these sources. The peak star formation rates are fairly uncertain due to a known degeneracy between the magnitude and length of the burst in this modeling \citep[see][]{Suess2022b}. 

We now seek to compare how those peak measurements compare between the CO luminous and non CO luminous post-starburst galaxies, and with literature samples at this earlier stage of evolution. In the leftmost panel of Figure \ref{fig:SFMS}, we show the full \squiggle CO-observed sample of galaxies with $\log(M_\star/M_\odot)>11$ relative to the star forming main sequence \citep[][at $z=0.7$]{Leja2022} at the inferred peak of their star formation. CO luminous (defined as $L'_{CO(2-1)}>2.5 \times 10^{9}$ K km s$^{-1}$ pc$^{-2}$, corresponding to $M_{H_2}>10^{10}$ $M_\odot$ for $\alpha_{CO}=4.0$) galaxies are shown in grey (with the three most CO-luminous galaxies highlighted in this program highlighted in purple), and CO non-luminous galaxies are shown in salmon. The median stellar mass and peak star formation rates of the two samples are consistent with each other, pointing to the plausibility of a common class of progenitors with peak 100 Myr averaged star formation rates of $\sim200$ $M_\odot$/yr. 

These high peak star formation rates point to clear classes of progenitor populations of co-eval, highly star forming, and infrared luminous galaxies \citep[e.g.,][]{Barro2016b, Chen2020_ULIRG}. We highlight a particular population of interest in the HizEA starburst and post-starburst sample \citep{Davis2023}. These galaxies have a number density at $z\sim0.5$ of $10^{-6}$ Mpc$^{-3}$, comparable to galaxies selected with the UBV \squiggle selections \citep{Setton2023}, and, crucially, the youngest have star formation rates of 100s of solar masses per year, either inferred by spectrophotometric fitting similar to the kind inferred in this work \citep[][Tremonti et al. in preparation]{Davis2023} or mid-IR/radio tracers \citep{Petter2020_IR_SFR}. In the left panel of Figure \ref{fig:SFMS}, we show the 10 Myr star formation rates of the HizEA galaxies with $t_\mathrm{light-weighted}<100$ Myr, consistent with being at their peak of star formation, in silver, showing that they comfortably span the same range in mass and star formation rate as our inferences for \squiggle post-starburst galaxies at their peak. In the central panel, we show the instantaneous star formation rates of the CO luminous \squiggle sample, along with the slightly older HizEA galaxies ($100<t_\mathrm{light-weighted}<200$ Myr) illustrating the decline from that peak activity that solidly places galaxies on the main sequence, and finally, in the last panel, we show our CO non-detected \squiggle galaxies and the oldest ($t_\mathrm{light-weighted}>200$ Myr) HizEA systems.

The morphological properties of the youngest HizEA compact starbursts offers a further point of connection between these galaxies and the \squiggle post-starbursts. These galaxies are compact ($r_e<1$ kpc) and highly disturbed, clearly at a late stage of a merger but often exhibiting multiple distinct nuclei with kpc-scale separations \citep{Sell2014, Diamond-Stanic2021}. The youngest, most CO luminous \squiggle galaxies also exhibit disturbed morphologies \citep{Spilker2022, Verrico2023, Donofrio2025}, but in what appears to be a later stage of mergers closer to coalescence. Similarly, the youngest \squiggle post-starburst galaxies are compact, with $r_e\sim2$ kpc, high \sersic indices, and central surface densities that are comparable to co-eval earlier-forming quiescent systems \citep{Setton2022}, consistent with other high-z massive post-starburst samples \citep[e.g.,][]{Almaini2017, Maltby2018, Wu2020, Zhang2024, Maltby2026_PSB_structure}. Additionally, the youngest HizEA-type galaxies host outflows with $\sim1000$ km/s velocities in their ISM \citep[][]{Tremonti2007,Sell2014,Geach2014,Geach2018,Perrotta2023_outflow}, which neatly connects to the idea, presented in Luo et al. 2026 submitted, that the $\sim200-400$ km/s outflows seen in \squiggle galaxies are, in part, relics that were launched during the peak starburst activity. These outflows, in conjunction with the dust-obscured star formation we observe here (and, potentially, a systematic change in the CO-to-$H_2$ conversion as the merger settles and star formation subsides), may connect the CO luminous post-starburst galaxies to their CO non-luminous post-starburst descendants, which we show in the third panel of Figure \ref{fig:SFMS}.

The consistency in number density (with the caveat that observability timescales remains a significant unknown), peak star formation rate, morphology, and outflow properties as interpreted in a direct evolutionary sequence paints a picture of the CO luminous post-starburst galaxies as the direct descendants of the compact, star forming, mid-merger young HizEA galaxies. Due to the degeneracy between burst length and magnitude, understanding the timescales associated with this decline from the peak star formation into the CO luminous post-starburst phase will require mass-complete spectroscopic samples at these redshifts to study the number density evolution of galaxies as a function of their star formation rates, structures, and stellar ages. The union of future spectroscopy surveys like the Prime Focus Spectrograph Galaxy Evolution Survey \citep{Greene2022} and Roman Space Telescope imaging for structural measurements will allow for much more quantitative timescales to be measured than \squigglecomma, which is limited by the incomplete targeting of SDSS at $z\sim0.5$ \citep{Suess2022a}.

\subsection{Are the CO luminous \squiggle galaxies ``post-starburst"?}

The star formation rates of post-starburst galaxies have been hotly debated for decades, primarily because while their rest-optical spectra suggest a burst that has concluded \citep[e.g.,][]{Dressler1983, Zabludoff1996, French2018a}, their IR and radio luminosities can often be interpreted as evidence for significant star formation in a dust-obscured phase \citep[e.g.,][]{Smail1999,Eales2018a,Eales2018b, Baron2022a, Baron2026_SED}, where galaxies still undergoing a peak of star formation can present in the optical as post-starburst due to dust geometry effects \citep[e.g.,][]{Cenci2025_PSB}. However, the IR luminosity is difficult to interpret as an indicator of the instantaneous star formation rate, especially in burst dominated systems where older stellar populations can contribute significantly to dust heating \citep{Hayward2014,Leja2018,Wild2025, Setton2025_squiggle}. Thus, while it is empirically clear that young post-starburst galaxies, both local \citep{French2015, Rowlands2015, Alatalo2016b, Smercina2018, Smercina2022, Luo2022, Baron2023} and high-z \citep{Suess2017, Belli2021, Bezanson2022a, Zanella2023, Setton2025_squiggle}, can have very high CO luminosities, the star forming state of that gas is a source of significant uncertainty in interpreting the evolutionary state of these systems.

The integrated emission line luminosities of dust-insensitive star formation rate tracers (e.g., Pa$\alpha$, [NeII]/[NeIII]) from near and mid-IR spectroscopy of local post-starburst systems point toward genuine suppression of star formation, with star formation rates from those lines that trend lower than the IR luminosity would predict, placing the galaxies below the main sequence \citep{Smercina2018, Luo2022}. In this work, we conduct the first resolved analysis of the near-IR recombination lines in the most CO luminous high-z systems, with the advantage of resolved JWST spectroscopy that enables dust corrections on $\sim700$ pc spatial scales. In the second panel of Figure \ref{fig:SFMS}, we show that the star formation rates that we measure--after correcting for the non-negligible nebular attenuation that is being experienced even at the wavelength of Pa$\alpha$--place these galaxies on or above the main sequence, in contrast with works based on rest-frame optical star formation rate indicators alone. 

The most meaningful result of this work is that even under the conservative assumption that ionization is dominated by current star formation, maximal star formation rates are 3-11$\times$ lower than the peak star formation rates experienced at earlier times. Thus, while we do find, consistent with the \cite{Setton2025_squiggle} fits that allowed for buried star formation, that post-starburst galaxies in the CO luminous phase are indeed forming stars, they are in a distinctly different phase than the peak of their starburst. During this period of dust obscured star formation, the star formation is centrally concentrated within $\sim2$ kpc and is less extended than the stellar continuum, indicating that the dust geometry does play a significant role in the inclusion of these galaxies in optical post-starburst selections, as seen in simulations \citep{Cenci2025_PSB}. But dust geometry alone cannot explain the full star formation histories of these sources; they must have experienced sustained periods of starburst in the period prior to observation, at rates that are in excess of their current instantaneous star formation rates, to form enough stars to produce their highly luminous rest-optical spectra. CO luminous post-starburst galaxies can still be accurately referred to as post-starburst, as, at the time of observation, they are securely in decline from that peak.

\section{Conclusions} \label{sec:conclusion}

In this work, we present new Magellan/FIRE and JWST NIRSpec G395M/170LP IFU spectroscopy of the three most CO-luminous post-starburst galaxies from the \squiggle Survey. Using this data, we make use of spatially-resolved Pa$\alpha$/Br$\gamma$/Br$\beta$ emission to map the dust attenuation and measure the global instantaneous star formation rates. We find evidence that Pa$\alpha$ is moderately attenuated in these galaxies, in excess of what would be predicted extrapolations of the H$\alpha$/H$\beta$ Balmer decrement. This suggests that there is deeply embedded star formation still occurring in the cores of these galaxies, cloaked by a high column of dust that entirely obscures the impact of these young stars in the rest-UV and optical. 

However, this residual star formation is still consistent with a significant decline from the peak star formation rate and cannot be modeled purely as a dust geometry effect in galaxies that are still at the peak of their starburst. Furthermore, sustained star formation at the observed level cannot clear out the molecular gas reservoirs inferred from CO on sub-Gyr timescales. This implies that the molecular gas reservoirs must be over-estimated, that there must be significant mechanical removal of gas, or that these galaxies are not the direct progenitors of the CO-non-luminous $\sim250$ Myr old post-starburst galaxies probed by the \squiggle Survey. While there is uncertainty about the future evolution of these galaxies, a strong ruling on their instantaneous star formation rate represents a strong step forward in our understanding of these galaxies that represent a crucial transitionary period of the rapid quenching process.

Our resolved dust attenuation and star formation rate measurements have only scratched the surface of the rich data contained in these JWST/NIRSpec IFU datacubes. The near-IR spectra of these galaxies contain a wealth of information about the molecular gas temperature and kinematics, the stellar populations that dominate at these wavelengths, and the neutral ISM. Future work will utilize this information, in conjunction with more-resolved CO emission maps and radio data, to constrain where these galaxies lie on the road to quiescence and to enhance our understanding of how merger-driven starbursts progress to produce the cores of massive, quiescent galaxies.

\facilities{JWST, ALMA, SDSS, Subaru, WISE, Herschel}

\software{Astropy \citep{astropy2013, astropy2018, astropy2022},
Matplotlib \citep{Hunter:2007}, Flexible Stellar Population Synthesis \citep{Conroy2009, Conroy2010}, SEDPy \citep{sedpy2019}, \texttt{Prospector} \citep{Johnson2017, Leja2017, Johnson2021}}

\begin{acknowledgements}

Support for this work was provided by The Brinson Foundation through a Brinson Prize Fellowship grant. Support for this work was provided by NASA through the NASA Hubble Fellowship grant \#HST-HF2-51618 awarded by the Space Telescope Science Institute, which is operated by the Association of Universities for Research in Astronomy, Incorporated, under NASA contract NAS5-26555. KAS, JSS, and VRD gratefully acknowledge support from NSF-AAG\#2407954 and 2407955, and NRAO-SOSPA11-006. DJS acknowledges helpful conversations with Taylor Hutchinson and Gourav Khullar regarding IFU reductions, Joel Leja regarding IR dust laws and \texttt{Prospector} treatment of energy balance, and Christy Tremonti for graciously sharing HizEA star formation history fits.

This work is based in part on observations made with the NASA/ESA/CSA James Webb Space Telescope. The data were obtained from the Mikulski Archive for Space Telescopes at the Space Telescope Science Institute, which is operated by the Association of Universities for Research in Astronomy, Inc., under NASA contract NAS 5-03127 for JWST. These observations are associated with program \#6719. This paper includes data gathered with the 6.5 meter Magellan Telescopes located at Las Campanas Observatory, Chile. 

This paper makes use of the following ALMA data: ADS/JAO.ALMA \#2016.1.01126.S, ADS/JAO.ALMA \#2017.1.01109.S, ADS/JAO.ALMA \#2021.1.01535.S, ADS/JAO.ALMA \#2021.1.00988.S,ADS/JAO.ALMA \#2021.1.00761.S, and ADS/JAO.ALMA \#2022.1.00604.S. ALMA is a partnership of ESO (representing its member states), NSF (USA) and NINS (Japan), together with NRC (Canada), NSTC and ASIAA (Taiwan), and KASI (Republic of Korea), in cooperation with the Republic of Chile. The Joint ALMA Observatory is operated by ESO, AUI/NRAO and NAOJ. The National Radio Astronomy Observatory is a facility of the National Science Foundation operated under cooperative agreement by Associated Universities, Inc. 

Funding for the Sloan Digital Sky Survey IV has been provided by the Alfred P. Sloan Foundation, the U.S. Department of Energy Office of Science, and the Participating Institutions. SDSS acknowledges support and resources from the Center for High-Performance Computing at the University of Utah. The SDSS web site is www.sdss4.org.

SDSS is managed by the Astrophysical Research Consortium for the Participating Institutions of the SDSS Collaboration including the Brazilian Participation Group, the Carnegie Institution for Science, Carnegie Mellon University, Center for Astrophysics | Harvard \& Smithsonian (CfA), the Chilean Participation Group, the French Participation Group, Instituto de Astrofísica de Canarias, The Johns Hopkins University, Kavli Institute for the Physics and Mathematics of the Universe (IPMU) / University of Tokyo, the Korean Participation Group, Lawrence Berkeley National Laboratory, Leibniz Institut für Astrophysik Potsdam (AIP), Max-Planck-Institut für Astronomie (MPIA Heidelberg), Max-Planck-Institut für Astrophysik (MPA Garching), Max-Planck-Institut für Extraterrestrische Physik (MPE), National Astronomical Observatories of China, New Mexico State University, New York University, University of Notre Dame, Observatório Nacional / MCTI, The Ohio State University, Pennsylvania State University, Shanghai Astronomical Observatory, United Kingdom Participation Group, Universidad Nacional Autónoma de México, University of Arizona, University of Colorado Boulder, University of Oxford, University of Portsmouth, University of Utah, University of Virginia, University of Washington, University of Wisconsin, Vanderbilt University, and Yale University.

The authors used Claude (Anthropic) to assist with code development and proofreading of this manuscript. All AI-assisted content was reviewed, verified, and edited by the authors, who take full responsibility for the accuracy and integrity of this work.

\end{acknowledgements}

\appendix
\restartappendixnumbering

\section{Data for J0910+0218 and J0907+0423}

\begin{figure*}
    \includegraphics[width=\textwidth]{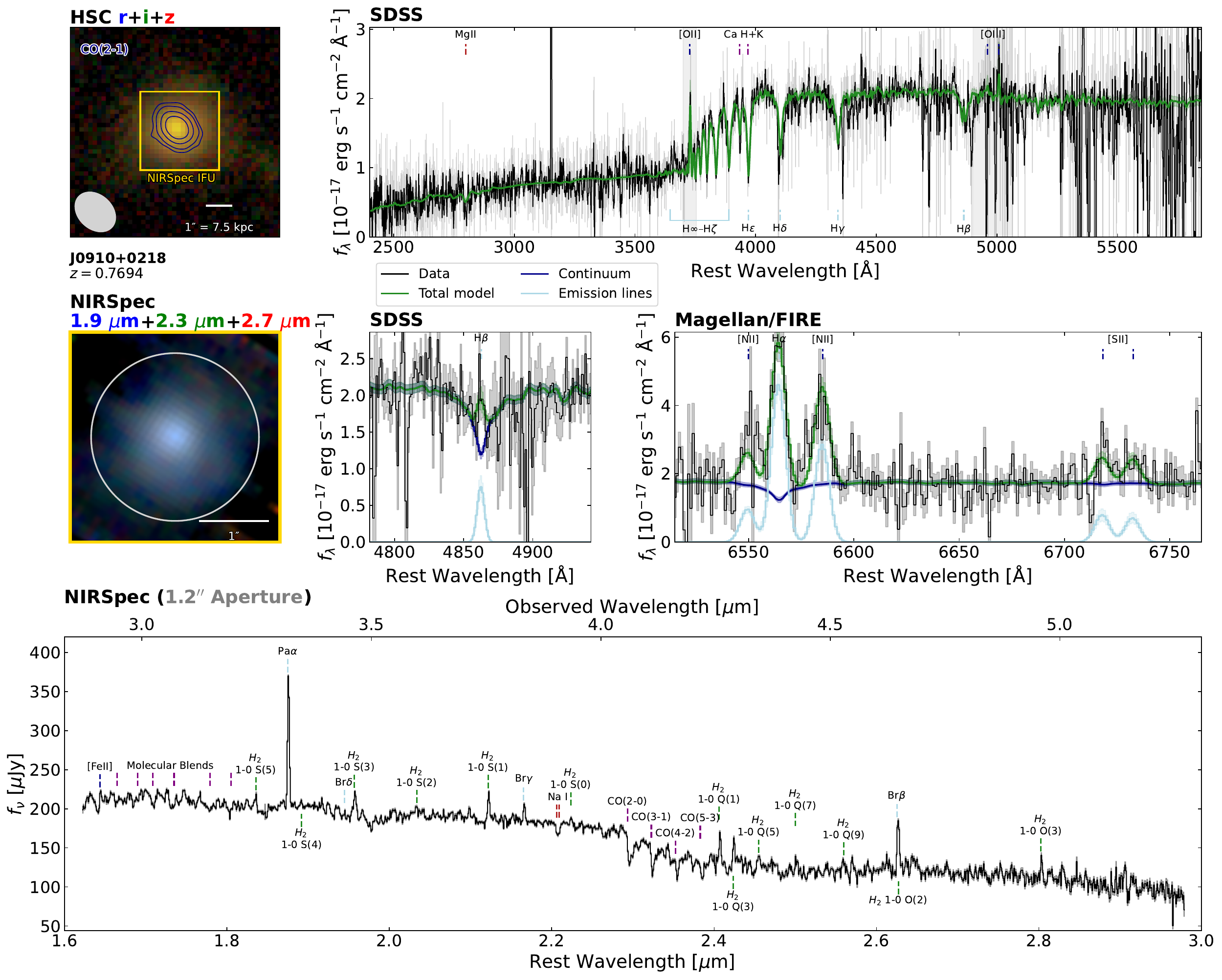}
    \caption{As in Figure \ref{fig:J1157_fulldata}, but showing J0910+0218.}
    \label{fig:J0910_fulldata}
\end{figure*}

\begin{figure*}
    \includegraphics[width=\textwidth]{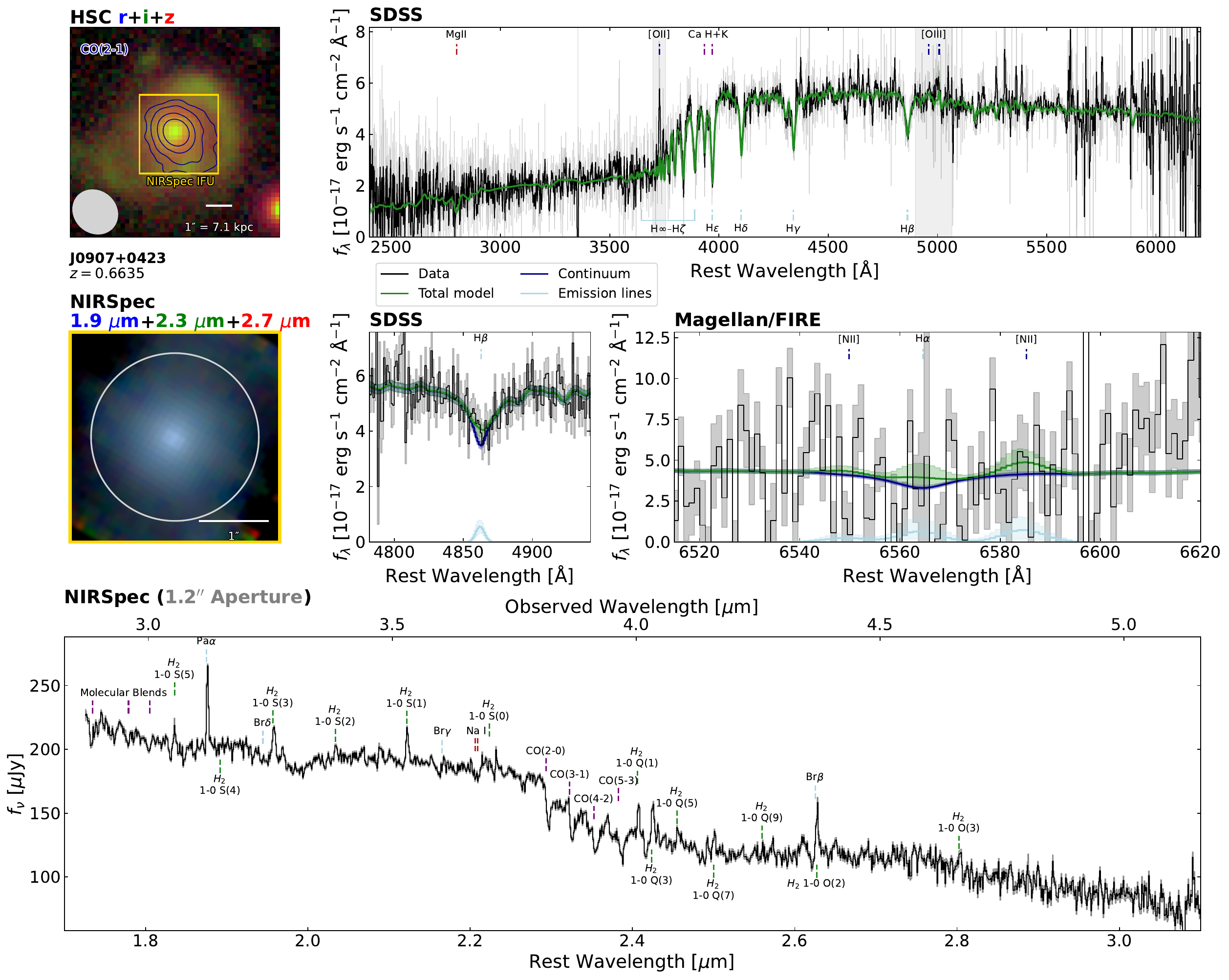}
    \caption{As in Figure \ref{fig:J1157_fulldata}, but showing J0907+0423.}
    \label{fig:J0907_fulldata}
\end{figure*}

In Figure \ref{fig:J0910_fulldata} and \ref{fig:J0907_fulldata}, we show (in the same format as in Figure \ref{fig:J1157_fulldata}) the Subaru/HSC, SDSS, Magellan/FIRE, and JWST/NIRSpec for J0910+0218 and J0907+0423.


\bibliography{SQuIGGLE-PaA}{}
\bibliographystyle{aasjournalv7}

\end{document}

%% file: tables/source_properties_table.tex
\begin{deluxetable*}{ccccccc}
\tablecaption{Host galaxy and molecular gas properties of the three CO-luminous \squiggle galaxies studied in this work. \label{tab:source_props}}
\tablewidth{0pt}
\tablehead{
    \colhead{ID} &
    \colhead{R.A.} &
    \colhead{Dec.} &
    \colhead{$z_\mathrm{spec}$} &
    \colhead{$\log(M_\star/M_\odot)$\tablenotemark{a}} &
    \colhead{$L'_{\mathrm{CO(2-1)}}$\tablenotemark{b}} &
    \colhead{$\log(M_{\mathrm{H_2}}/M_\odot)$\tablenotemark{c}} \\
    \colhead{} &
    \colhead{[deg]} &
    \colhead{[deg]} &
    \colhead{} &
    \colhead{} &
    \colhead{[$10^{10}$ K km s$^{-1}$ pc$^{2}$]} &
    \colhead{}
    }
\startdata
    J1157+0132 & 179.48879 & 1.53753 & 0.7559 & 11.40$\pm^{0.04}_{0.05}$ & 2.83 $\pm$ 0.04 & 11.05 $\pm$ 0.01 \\
    J0910+0218 & 137.61906 & 2.30931 & 0.7694 & 11.12$\pm^{0.06}_{0.05}$ & 1.53 $\pm$ 0.04 & 10.79 $\pm$ 0.01 \\
    J0907+0423 & 136.99116 & 4.38426 & 0.6635 & 11.40$\pm^{0.04}_{0.08}$ & 1.44 $\pm$ 0.02 & 10.76 $\pm$ 0.01 \\
            \enddata
\tablenotetext{a}{From the $\tau_{BC}$-free spectrophotometric fits of \citet{Setton2025_squiggle}.}
\tablenotetext{b}{As measured in \citet{Setton2025_squiggle}.}
\tablenotetext{c}{Assuming $\alpha_{CO}=4.0\,M_\odot\,(\mathrm{K\,km\,s^{-1}\,pc^{2}})^{-1}$ and $r_{21}=1.0$.}
\end{deluxetable*}

%% file: tables/halpha_fit_table_95.tex
\begin{deluxetable*}{lcccccccc}
\tablecaption{Line fitting results from the fits to the rest-optical SDSS and Magellan/FIRE spectroscopy. All presented line fluxes are corrected by the aperture corrections ($c_\mathrm{ap}$) shown. Due to the uncertainty in the $H_\alpha/H_\beta$ ratio imparted by the very low significance $H_\beta$ corrections, we present these results with 95\% confidence intervals for all quantities to illustrate the broadness of the $A_V$ and SFR posteriors that rely on the measurement of that ratio.\label{tab:halpha}}
\tablewidth{0pt}
\tablehead{\colhead{ID} & \colhead{$F(\mathrm{H}\beta)$} & \colhead{$c_{\rm ap}(\mathrm{H}\beta)$} & \colhead{$F(\mathrm{H}\alpha)$} & \colhead{$c_{\rm ap}(\mathrm{H}\alpha)$} & \colhead{$\mathrm{[N\,II]}/\mathrm{H}\alpha$} & \colhead{$\sigma_v$} & \colhead{$A_V$} & \colhead{SFR} \\
           \colhead{} & \colhead{[$10^{-16}\,\mathrm{erg\,s^{-1}\,cm^{-2}}$]} & \colhead{} & \colhead{[$10^{-16}\,\mathrm{erg\,s^{-1}\,cm^{-2}}$]} & \colhead{} & \colhead{} & \colhead{[km\,s$^{-1}$]} & \colhead{[mag]} & \colhead{[$M_\odot\,\mathrm{yr^{-1}}$]}}
\startdata
J1157+0132 & $1.76^{+1.00}_{-0.93}$ & $2.15^{+0.14}_{-0.18}$ & $7.4^{+1.4}_{-1.3}$ & $2.80^{+0.23}_{-0.23}$ & $1.04^{+0.25}_{-0.21}$ & $293^{+92}_{-37}$ & $1.2^{+2.3}_{-1.2}$ & $27^{+130}_{-17}$ \\
J0910+0218 & $0.88^{+0.35}_{-0.35}$ & $1.67^{+0.11}_{-0.12}$ & $6.96^{+0.91}_{-0.89}$ & $2.41^{+0.19}_{-0.18}$ & $0.62^{+0.12}_{-0.11}$ & $154^{+21}_{-16}$ & $3.2^{+1.6}_{-1.1}$ & $117^{+280}_{-67}$ \\
J0907+0423 & $<1.4$ & $2.88^{+0.17}_{-0.17}$ & $<4.5$ & $4.34^{+0.51}_{-0.43}$ & \nodata & $186^{+56}_{-43}$ & $<5.3$ & $<170$
\enddata
\end{deluxetable*}

%% file: tables/nir_lines_table.tex
\begin{deluxetable}{lc}
\tablecaption{Near-infrared emission lines. \label{tab:nir_lines}}
\tablewidth{0pt}
\tablehead{
    \colhead{Line} &
    \colhead{Wavelength} \\
    \colhead{} &
    \colhead{($\mu$m)}
}
\startdata
\cutinhead{Hydrogen/Helium recombination lines}
    He\,\textsc{i}        & 1.8686 \\
    Pa$\alpha$            & 1.8751 \\
    Br$\delta$            & 1.9446 \\
    He\,\textsc{i}        & 2.0587 \\
    Br$\gamma$            & 2.1661 \\
    Br$\beta$             & 2.6259 \\
\cutinhead{Molecular hydrogen lines}
    H$_2$ 1--0 S(5)       & 1.8358 \\
    H$_2$ 1--0 S(4)       & 1.8920 \\
    H$_2$ 1--0 S(3)       & 1.9576 \\
    H$_2$ 1--0 S(2)       & 2.0338 \\
    H$_2$ 1--0 S(1)       & 2.1218 \\
    H$_2$ 1--0 S(0)       & 2.2235 \\
    H$_2$ 2--1 S(1)       & 2.2477 \\
    H$_2$ 1--0 O(2)       & 2.6269 \\
    H$_2$ 1--0 O(3)       & 2.8025 \\
\cutinhead{Other lines}
    {[Fe\,\textsc{ii}]}   & 1.6440 \\
\enddata
\tablecomments{}
\end{deluxetable}

%% file: tables/paa_sfr_table.tex
\begin{deluxetable*}{ccccccc}
\tablecaption{NIR emission line-derived dust corrections and star formation rates. \label{tab:paa_sfr}}
\tablewidth{0pt}
\tablehead{
    \colhead{ID} &
    \colhead{$L_{\mathrm{Pa}\alpha,\,\mathrm{obs}}$} &
    \colhead{$L_{\mathrm{Pa}\alpha,\,\mathrm{corr}}$} &
    \colhead{$A_{\mathrm{Pa}\alpha,\,\mathrm{eff}}$\tablenotemark{a}}  & \colhead{$A_{V,\,\mathrm{eff}}$\tablenotemark{b}} &
    \colhead{SFR$_{\mathrm{Pa}\alpha}$\tablenotemark{c}} &
    \colhead{log(sSFR$_{\mathrm{Pa}\alpha}$)\tablenotemark{c}} \\
    \colhead{} &
    \colhead{[$10^{41}$ erg s$^{-1}$]} &
    \colhead{[$10^{41}$ erg s$^{-1}$]} &
    \colhead{[mag]} &
    \colhead{[mag]} &
    \colhead{[$M_\odot\,\mathrm{yr}^{-1}$]} &
    \colhead{[log($\mathrm{yr^{-1}}$)]}
    }
\startdata
    J1157+0132 & 4.90$\pm ^{0.07} _{0.07}$  & 14.1$\pm ^{2.6} _{1.9}$  & 1.15$\pm ^{0.18} _{0.16}$  & 11.3$\pm ^{1.8} _{1.6}$  & 65.0$\pm ^{12.0} _{8.8}$ (37.7$\pm ^{6.9} _{5.1}$)  & -9.59$\pm ^{0.09} _{0.07}$ (-9.82$\pm ^{0.09} _{0.07}$)  \\
    J0910+0218 & 6.38$\pm ^{0.05} _{0.05}$  & 17.4$\pm ^{1.9} _{1.8}$  & 1.09$\pm ^{0.11} _{0.12}$  & 10.8$\pm ^{1.1} _{1.2}$  & 79.9$\pm ^{8.9} _{8.2}$ (54.5$\pm ^{6.1} _{5.6}$)  & -9.22$\pm ^{0.07} _{0.08}$ (-9.38$\pm ^{0.07} _{0.08}$)  \\
    J0907+0423 & 2.21$\pm ^{0.06} _{0.05}$  & 5.80$\pm ^{1.20} _{0.86}$  & 1.04$\pm ^{0.20} _{0.17}$  & 10.3$\pm ^{2.0} _{1.7}$  & 26.7$\pm ^{5.4} _{3.9}$ (16.8$\pm ^{3.4} _{2.5}$)  & -9.97$\pm ^{0.09} _{0.09}$ (-10.17$\pm ^{0.09} _{0.09}$)  \\
            \enddata
\tablenotetext{a}{Assuming a power-law near-IR dust law, with $\alpha=2.11$ \citep{Fritz2011_nearIR_dust}.}
\tablenotetext{b}{Converted from $A_{\mathrm{Pa}\alpha,\,\mathrm{eff}}$ using the joint optical+NIR extinction curve of \citet{Wang2019_dust_law}, which has a near-identical near-IR parameterization to \cite{Fritz2011_nearIR_dust}.}
\tablenotetext{c}{Value in parentheses is corrected for the maximal contribution from non-star formation ionization to Pa$\alpha$ under the mixing model of Sec.~\ref{subsec:non-sf-ionization}: $([\mathrm{FeII}]/\mathrm{Pa}\alpha)_\mathrm{non-SF}$ set to the highest securely detected (SNR $\geq 5$ for both lines) dust-corrected [FeII]/Pa$\alpha$ spaxel in each galaxy, giving integrated $f_\mathrm{non-SF}$ of 42.0\% and 31.8\% for J1157+0132 and J0910+0218 respectively. J0907+0423 lacks [FeII] coverage and adopts the median of these two galaxies' values.}
\end{deluxetable*}